# Free-viewpoint Indoor Neural Relighting from Multi-view Stereo

JULIEN PHILIP, Inria, Université Côte d'Azur and Adobe Research
SÉBASTIEN MORGENTHALER, Inria, Université Côte d'Azur
MICHAËL GHARBI, Adobe Research
GEORGE DRETTAKIS, Inria, Université Côte d'Azur

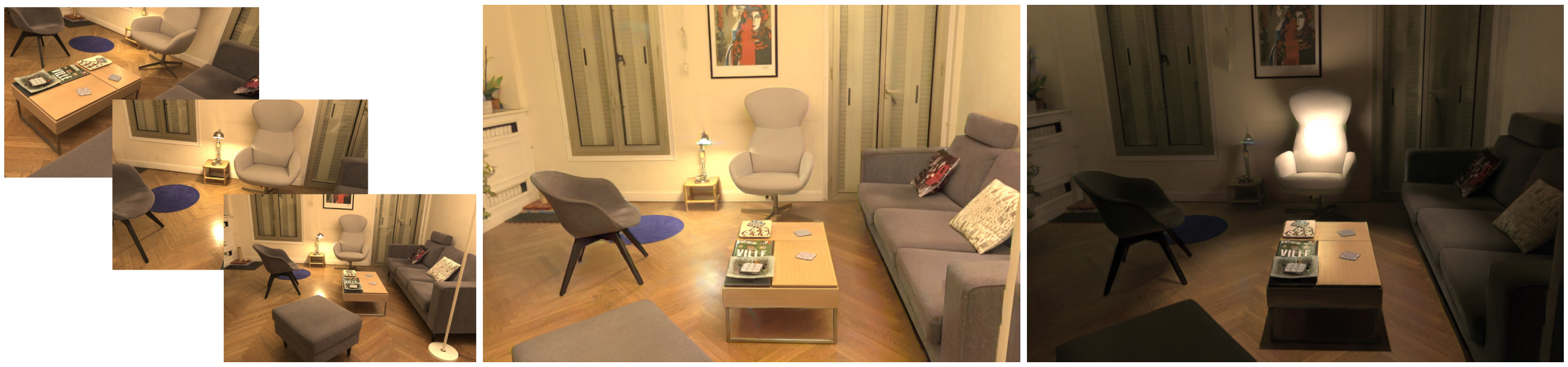

(a) three of $n$ input views (b) novel view synthesized by our algorithm (c) the same novel view with modified illumination

Fig. 1. Our neural algorithm takes as input a multi-view capture of a real scene (250-350 RAW photos) under a single lighting condition (a). It re-renders the scene from novel viewpoints, accounting for view-dependent glossy effects, e.g., on the floor (b) and on the table (c), allowing free-viewpoint navigation. Our network internally builds an implicit representation of the scene materials, based on complex input feature maps we compute with a novel hybrid image- and physically-based rendering algorithm. This enables a user to insert new lights in the scene and/or turn off the original illumination; the network produces a new rendering of the scene under the modified lighting (c). Note how the light – that was on in the input photos – has been turned off with our method.

We introduce a *neural relighting* algorithm for captured indoors scenes, that allows interactive *free-viewpoint* navigation. Our method allows illumination to be changed synthetically, while coherently rendering cast shadows and complex glossy materials. We start with multiple images of the scene and a 3D mesh obtained by multi-view stereo (MVS) reconstruction. We assume that lighting is well-explained as the sum of a view-independent diffuse component and a view-dependent glossy term concentrated around the mirror reflection direction. We design a convolutional network around input feature maps that facilitate learning of an implicit representation of scene materials and illumination, enabling both relighting and free-viewpoint navigation. We generate these input maps by exploiting the best elements of both image-based and physically-based rendering. We sample the input views to estimate diffuse scene irradiance, and compute the new illumination caused by user-specified light sources using path tracing. To facilitate the network's understanding of materials and synthesize plausible glossy reflections, we reproject the views and compute *mirror images*. We train the network on a synthetic dataset where each scene is also reconstructed with MVS. We show results of our algorithm relighting real indoor scenes and performing free-viewpoint navigation with complex and realistic glossy reflections, which so far remained out of reach for view-synthesis techniques.

Authors' addresses: Julien Philip, Inria, Université Côte d'Azur and Adobe Research; Sébastien Morgenthaler, Inria, Université Côte d'Azur; Michaël Gharbi, Adobe Research; George Drettakis, Inria, Université Côte d'Azur.

## 1 INTRODUCTION

A key challenge for virtual reality is the ability to easily capture, edit and re-render indoor scenes from any viewpoint. Controlling the scene illumination, in particular, is key to many immersive applications like architectural pre-visualization or game content-creation. State-of-the-art image-based rendering methods [Hedman et al. 2018; Mildenhall et al. 2020] offer no relighting control, and they struggle to render the complex glossy reflections that are ubiquitous in man-made environments. Likewise, recent relighting techniques rarely account for view-dependent lighting effects [Wu and Saito 2017]. Previous methods usually involve complex setups to acquire the geometry, materials and illumination [Debevec 2006; Tchou et al. 2004]), and use expensive inverse rendering models [Yu et al. 1999].

We introduce an efficient neural renderer for free-viewpoint navigation that can relight *indoor* scenes. Indoors environments often contain glossy surfaces, which leads to complex view-dependent effects. Our algorithm can relight scenes with such materials. It uses a convolutional network to extract an implicit representation of the scene's materials and lighting from a multi-view dataset captured under a *single* lighting condition with a wide baseline, and synthesize relit novel views. We ask the user to specify a target illumination by placing virtual lights sources in a 3D reconstruction of the scene. Except for a short pre-computation of the novel illumination, our renderer runs at interactive frame rates.

Our algorithm combines elements of physically-based path-tracing to simulate the new illumination, and Image-Based Rendering (IBR) to reproject and blend data from the input views to the new view. Unlike pure image-based techniques, our solution can change the lighting of scenes and handles complex glossy materials, which are abundant indoors. Compared to inverse rendering methods, it is faster, requires no explicit material model, and is robust to imperfect geometry given a reasonable multi-view stereo reconstruction.

Our inputs are wide-baseline multi-view captures — typically 250–350 RAW photos — from which we compute a 3D mesh by Multi-View Stereo [Furukawa and Ponce 2010; Jancosek and Pajdla 2011; RealityCapture 2019]. So, compared to single-image methods [Wu and Saito 2017], we start with richer information. Since our network is convolutional, we summarize this information into 2D input feature maps that are reprojected in the frame of the novel view. We make the simplifying assumption that the lighting can be decomposed into the sum of two components: view-independent diffuse, and view-dependent glossy, and we design our feature maps to best characterize scene materials, source, and target lighting conditions under this approximation. Specifically, our input features maps rely on three main elements.

First, we encode the diffuse illumination with irradiance maps. We start by estimating the source diffuse illumination via a lightweight Monte Carlo integration and store it into an irradiance map for each input viewpoint. From this we compute an approximate albedo, stored at mesh vertices. Albeit a crude approximation, the albedo mesh is enough to compute approximate diffuse irradiance maps for the user-specified target illumination, also using path-tracing. Analyzing reprojections of the source and the target irradiance maps, our network learns to change the diffuse illumination, including overall energy levels, cast shadows and global illumination.

Second, to deal with the notoriously difficult case of glossy materials, we guide the analysis and synthesis of view-dependent effects using *mirror images*. We compute these using a fast single-ray mirror reflection calculation for both source and target lighting conditions. Source mirror images help our model analyze scene materials and discriminate between diffuse and view-dependent image components. They can be precomputed once per scene and stored at the input viewpoints. The target mirror image is generated on the fly, at render time, guiding the synthesis of new glossy reflections.

Third, we reproject the input images and corresponding feature maps (i.e., originally in the space of each input camera) using image-based rendering techniques so that the network can render arbitrary viewpoints at interactive frame rates. Our reprojections are carefully designed to give the network a diverse set of observations for each surface in the scene, which helps characterize scene material properties like roughness.

We train our network entirely on synthetic data. To ensure the model generalizes well to real scenes, we use Multi-View Stereo (MVS) reconstructions of the synthetic scenes instead of the ground-truth geometry to compute our input features at training time [Philip et al. 2019].

We demonstrate our method on a variety of real indoor scenes (living rooms, kitchen, bedrooms, hallway), showing interactive sessions in which a user adds lights, turns off the original light sources, and freely navigates around the scene. We compare our approach quantitatively against view-synthesis techniques on a held-out synthetic scene.

In short, we make the following contributions:

- The first algorithm to enable relighting and interactive free-viewpoint navigation of indoor scenes, with realistic rendering of glossy materials.
- A new hybrid rendering pipeline that computes input feature buffers characterizing the scene illumination using physically-based rendering to guide a neural relighting model.
- A new image-based rendering reprojection scheme that pools together multiple material hypotheses from the input views, including pure *mirror images*, to help the neural renderer analyze scene materials (source mirror images) and synthesize realistic glossy reflections (target mirror images).

Our interactive relighting system source code, training scripts, models, test scenes, and training images are available at https://repo-sam.inria.fr/fungraph/deep-indoor-relight/.

## 2 PREVIOUS WORK

Our algorithm builds on a wealth of past research; we only review the closest work in material estimation, relighting and rendering.

### 2.1 Material estimation

Intrinsic image decompositions [Tappen et al. 2003] assume images are the product of diffuse reflectance and shading, with impressive success [Bonneel et al. 2017]. The scenes we consider contain glossy materials that violate this assumption.

Many solutions exist to estimate spatially-varying bi-directional distribution functions (SVBRDFs). Early optimization-based methods are successful for individual objects [Lensch et al. 2003] but require constraining capture conditions. Elaborate hardware like the Light Stage [Debevec et al. 2000] use multiple lights and/or cameras to record highly detailed representations of complex materials like the human skin. Under certain assumptions like repetitive texture, lighter-weight methods, e.g., based on flash/no flash photos [Aittala et al. 2015], can extract complex SVBRDFs. Neural material estimation [Deschaintre et al. 2018; Li et al. 2017] enables one-shot SVBRDF estimation; they are typically trained on synthetic data, sometimes using differentiable renderers [Azinovic et al. 2019; Li et al. 2018a; Nimier-David et al. 2019]. Recent methods even handle full objects [Li et al. 2018b]. Li et al. [2019] and Sengupta et al. [2019] estimate scene albedo, normals and lighting from a single image, but do not handle complex materials. Our method is not limited to isolated objects with diffuse materials. It can analyze materials by observing the scene from multiple viewpoints, and handle complete scenes with glossy surfaces without explicitly estimating material properties. Meka et al. [2018] recover sharp reflections with a network trained to produce *mirror images*; they are limited to individual segmented objects. Our mirror images are related but we provide them as inputs rather than training targets. They guide material understanding and glossy highlight synthesis.

### 2.2 Light estimation and image relighting

Most current light estimation techniques work on single images and focus on compositing virtual objects in a real image, rather than significantly modifying the scene's illumination [Gardner et al. 2019, 2017; Hold-Geoffroy et al. 2017; LeGendre et al. 2019; Murmann et al. 2019]. Several methods have been proposed to relight photos [Wu and Saito 2017], many of them tailored to human faces [Marschner and Greenberg 1997], recently with deep learning approaches [Meka et al. 2019; Sun et al. 2019]. Deep learning also powers object relighting techniques that use multiple lighting conditions as input [Xu et al. 2018]. Although they provide many interesting intuitions, these methods focus on single images, which means they are not directly compatible with our goal of free-viewpoint 3D navigation. We target multi-view datasets captured with a commodity camera. Earlier works needed user-assisted geometry [Yu et al. 1999], or multiple input lighting setups [Loscos et al. 1999]; these solutions typically used global illumination for inverse rendering. Alternative methods such as Precomputed Radiance Transfer [Sloan et al. 2002] can also be used in such contexts. Given a set of RGBD images of a furnished indoor scene, Zhang et al. [2016] recover a 3D model of the empty room they can relight. However, their method cannot handle the original, furnished scene. Wei et al. [2020] tackle the problem of light estimation of an object captured by an RGBD camera. Similar to our solution they use an irradiance map and mirror reflections for specular estimation. However, these are coarse estimates based on distant lighting – i.e., environment maps – and cannot handle the full scenes with local lighting that we target; this is also the case for the method of Song and Funkhouser [2019].

Laffont et al. [2012] and Duchêne et al. [2015] use multi-view intrinsic images for relighting, from several or a single input lighting condition, respectively. Recently, Philip et al. [2019] used deep learning to relight outdoor scenes captured with a single lighting condition. They compute RGB shadow images from MVS geometry which their network refines to synthesize convincing cast shadows. However, their algorithm assumes a distant sun/sky model which is too limiting for general indoor scenes. [Guo et al. 2019; Zhang et al. 2021] focus on humans. They capture data in a light stage whereas we acquire unstructured images under a single illumination. Xu et al. [2019] can relight single objects from multi-view captures but their acquisition setup requires multiple illumination conditions. Neural re-rendering [Meshry et al. 2019] also takes varying lighting as input using internet images, allowing transitions between different conditions. In contrast, we target relighting and rendering of scenes with strong indirect and glossy illumination, such as indoor environments, captured under a single illumination condition.

### 2.3 Image-based and neural rendering

Image-based rendering (IBR) methods blend input images to synthesize a novel view [Buehler et al. 2001], often using reconstructed multi-view stereo (MVS) geometry for free-viewpoint navigation. Recently, per-view [Chaurasia et al. 2013; Hedman et al. 2016] or volumetric [Penner and Zhang 2017] data has been used to improve errors due to inaccuracies of MVS reconstruction. These methods typically cannot alter the scene illumination after capture, and because they blend photos, they cannot render glossy materials: highlights have a view-dependent flow, different from camera motion. Deep learning has been applied to view synthesis. Early solutions include appearance flow [Zhou et al. 2016] which used a CNN to reproject views and DeepStereo [Flynn et al. 2016] that learns a plane sweep for novel view synthesis. Deep Blending learns blending weights for image mosaics generated by improved per-view meshes [Hedman et al. 2018]. To synthesize novel views, Riegler and Koltun [2020] aggregate features from reprojected images using a recurrent network. Multi-plane images have been used in the context of small-baseline light fields [Flynn et al. 2019; Mildenhall et al. 2019; Zhou et al. 2018]. Another neural representation, deep neural textures [Thies et al. 2019] allows view synthesis for glossy objects at the cost of very dense capture. Chen et al. [2020] use them to relight single objects. More recent techniques include voxel-based [Lombardi et al. 2019; Sitzmann et al. 2019a], and implicit deep representations [Bemana et al. 2020; Mildenhall et al. 2020; Niemeyer et al. 2020; Sitzmann et al. 2019b]. Despite impressive advances, our comparisons show neural rendering still struggles with the large baseline, single lighting setting we work with, both in terms of glossy effects fidelity and for free-viewpoint navigation more generally.

## 3 MULTI-VIEW NEURAL RELIGHTING

We start from a set of photographs and a 3D mesh reconstructed by Multi-View Stereo. Our goal is to relight the scene — by adding new lights in the virtual 3D space or disabling the original illumination — and enable free-viewpoint navigation. As the user navigates in 3D, our algorithm synthesizes realistic novel views with the modified lighting. We pay special attention to the synthesis of plausible glossy reflections, which are key to realism, especially in indoor scenes.

Relighting requires an estimate of the *source* (i.e., at the time of capture) and of the *target* illumination. To facilitate the source estimation, we require that the original lights be visible in the input photos. Instead of explicitly modeling the lights and materials — which is a difficult inverse problem that requires costly physically based rendering — we train a convolutional neural network to learn an implicit representation and synthesize the relit novel view.

Our design is motivated by a simplified illumination model (§ 3.1). We encode the source and target illumination in easy-to-compute features maps from which the network predicts the final output. These maps are produced by a novel hybrid renderer that blends elements of physically and image-based rendering (§ 4).

### 3.1 Simplified illumination model

The rendering equation for a static, non-emissive surface is:

$$L_o(x,\omega_o) = \int f(x,\omega_i,\omega_o) L_i(x,\omega_i)\omega_i \cdot n d\omega_i, \tag{1}$$

where $L_o$ and $L_i$ are the outgoing and incoming radiance, in direction $\omega_o$ and $\omega_i$ respectively, at a 3D point $x$ with normal $n$. We approximate the bidirectional reflectance distribution function $f$ as the sum of two components:

$$f(x,\omega_i,\omega_o) \approx a(x) + s(x,\omega_i,\omega_o), \tag{2}$$

where $a(x)$ is a diffuse albedo, independent of viewing angles, and $s(x,\omega_i,\omega_o)$ is a view-dependent glossy term, compactly supported in a narrow angular neighborhood of the mirror direction at $x$. This approximation is reasonable for isotropic materials. Integrating over incident angles gives the following decomposition of the outgoing radiance:

$$L_o(x,\omega_o) \approx a(x)E(x) + S(x,\omega_o), \tag{3}$$

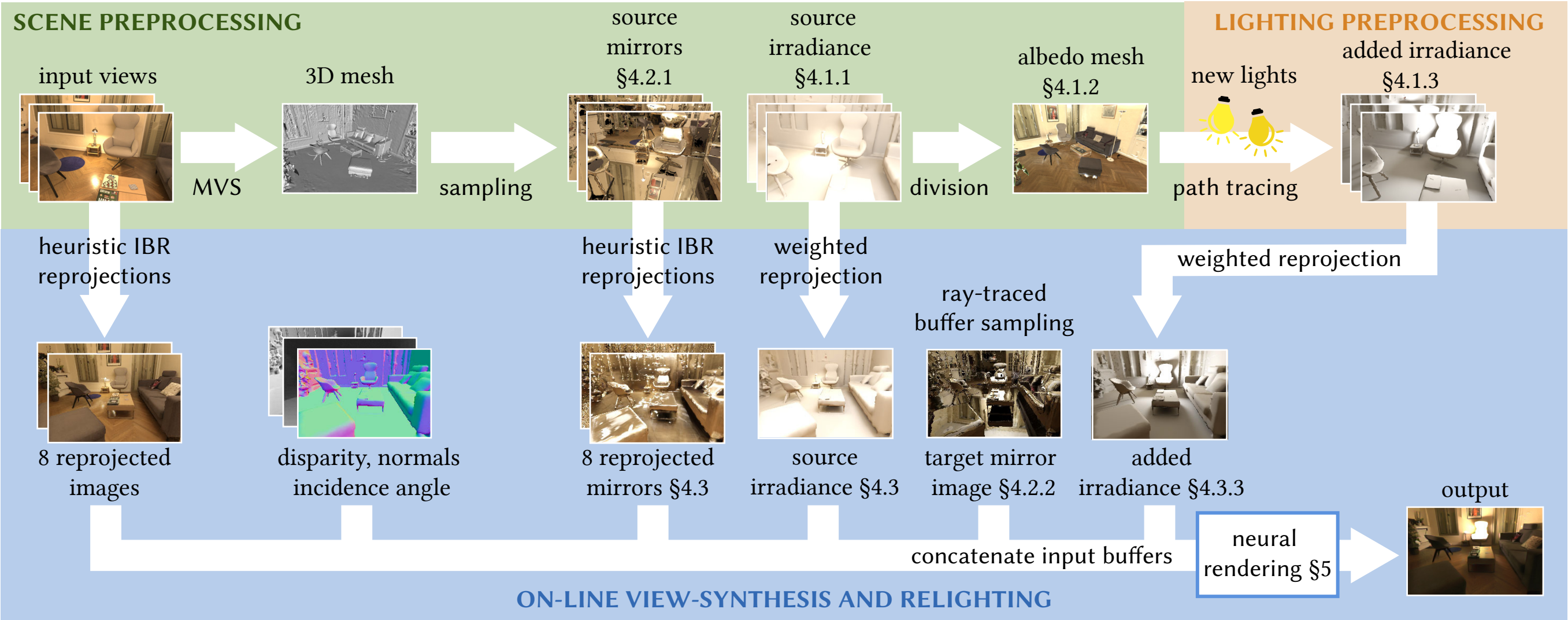

Fig. 2. Overview of our pipeline. We first perform scene pre-processing of our multi-view dataset and MVS geometry to precompute source diffuse irradiance, source mirror images and an albedo mesh, stored with the scene. For a new lighting condition target added irradiance is precomputed. Online view synthesis is achieved by reprojecting the information into a novel view using our neural network to synthesize a relit image.

with $E(x) = \int L_i(x, \omega_i)\omega_i \cdot n d\omega_i$ the diffuse irradiance, and $S(x, \omega_o) = \int s(x, \omega_i, \omega_o) L_i(x, \omega_i)\omega_i \cdot n d\omega_i$ the near-mirror glossy illumination. Our strategy is to design a hybrid rendering algorithm to estimate $E$ and $S$, using purely diffuse and purely mirror guide maps, from which a neural network can learn an implicit representation of the scene for use in relighting and view-synthesis.

### 3.2 Encoding scene illumination as 2D feature maps

Image relighting is extremely ill-posed; learning complex light transport interactions between distant objects with a standard convolutional network is difficult. To make the problem tractable, we estimate $E$ and $S$ and encode them as a set of 2D input maps, which the network implicitly uses to decouple intrinsic material properties and illumination. Maps that describe the source illumination conditions are precomputed once and stored with the scene. Maps characterizing the novel illumination require a short preprocessing step (a few minutes), computed once for each additional user-defined light. A final step reprojects the precomputed per-view 2D maps and blends them to produce 2D composites in the new view. Section 4 describes our hybrid rendering algorithm to compute this representation. We then process the composites with a neural network (§ 5) that synthesizes novel views under the modified lighting. The reprojection and neural rendering phases run at interactive frame rates.

## 4 GENERATING THE NETWORK INPUTS

We build our 2D representation around three main components. First, to characterize the lighting, we precompute *diffuse irradiance maps* for each view by stochastic integration, sampling the input images (§ 4.1). From these maps we build an albedo mesh, which we then use to precompute another set of irradiance maps describing the energy added by each user-defined light.

Second, we compute *mirror images* that encode the radiance in the mirror direction at each surface point (§ 4.2). We do this for both source and target illuminations. Source mirror images are precomputed once for each input view. The target mirror is computed dynamically, for the novel view only.

Third, we dynamically *reproject and blend* the input images (resp. mirror images, irradiance maps) in the novel view to produce a fixed number of composite feature maps that provide rich cues about real-world light transport and encode the illumination changes requested by the user (§ 4.3).

The reprojected composite features are concatenated and passed to the neural network which predicts the relit novel view (§ 5). Figure 2 summarizes our rendering pipeline.

### 4.1 Irradiance maps as illumination descriptors

Source irradiance maps $E^{src}_{i..n}$ describe the lighting of the input scene (§ 4.1.1). They are stored in the original cameras' reference frames. Dividing the input images $I_{i..n}$ by their corresponding irradiance maps, we construct a diffuse albedo mesh (§ 4.1.2) that — albeit a coarse approximation — is sufficient to estimate the irradiance added by user-specified lights. We compute the additional illumination using path-tracing on the colored mesh and store it in per-view maps $E^{add}_{i..n}$. These maps are precomputed for each new light (§ 4.1.3). We also let the user dim (or turn off) the original lighting in the scene (§ 4.1.4). The energy to be removed is encoded in the maps $E^{rem}_{i..n}$.

*4.1.1 Diffuse source irradiance - $E^{src}_{1..n}$.* We estimate a source irradiance map for each input view by stochastic integration. For each pixel in each view, we cast a ray in the 3D scene. If the ray intersects the geometry, we cosine importance sample 128 secondary rays in the hemisphere above the intersection, integrating the pixel colors from the remaining images, which are stored as 16 bit floats; please see Fig. 3. In Appendix A, we explain how we estimate irradiance in the presence of overexposed pixels.

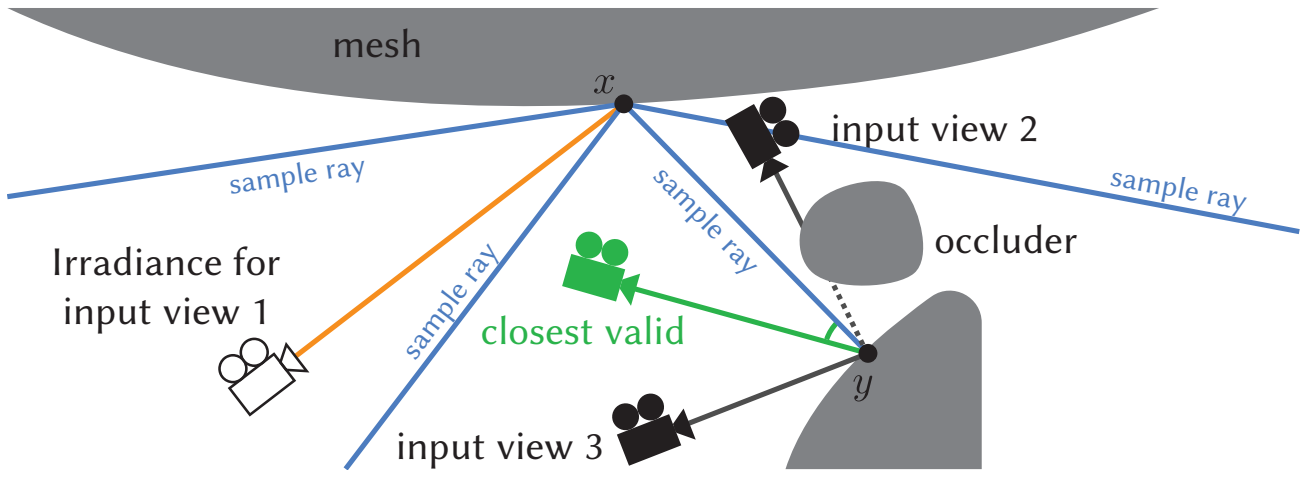

Fig. 3. To compute the diffuse source irradiance for input view 1, for each point $x$ corresponding to a pixel we cosine importance sample 128 directions, and integrate the color from the input view with the smallest angle with the ray. We ignore cameras that do not see the intersection $y$ between the sample ray and the geometry using a visibility test.

Specifically, each secondary ray samples its color from the image whose camera-intersection vector forms the smallest angle with the ray, which takes into account light directionality and minimizes reprojection errors in case of poor reconstruction. We ignore secondary rays that do not reproject to any of the input views. We denoise the irradiance maps using Optix [Chaitanya et al. 2017] to avoid magnifying the noise from the stochastic integration when reprojecting. Figure 4b shows a source irradiance map.

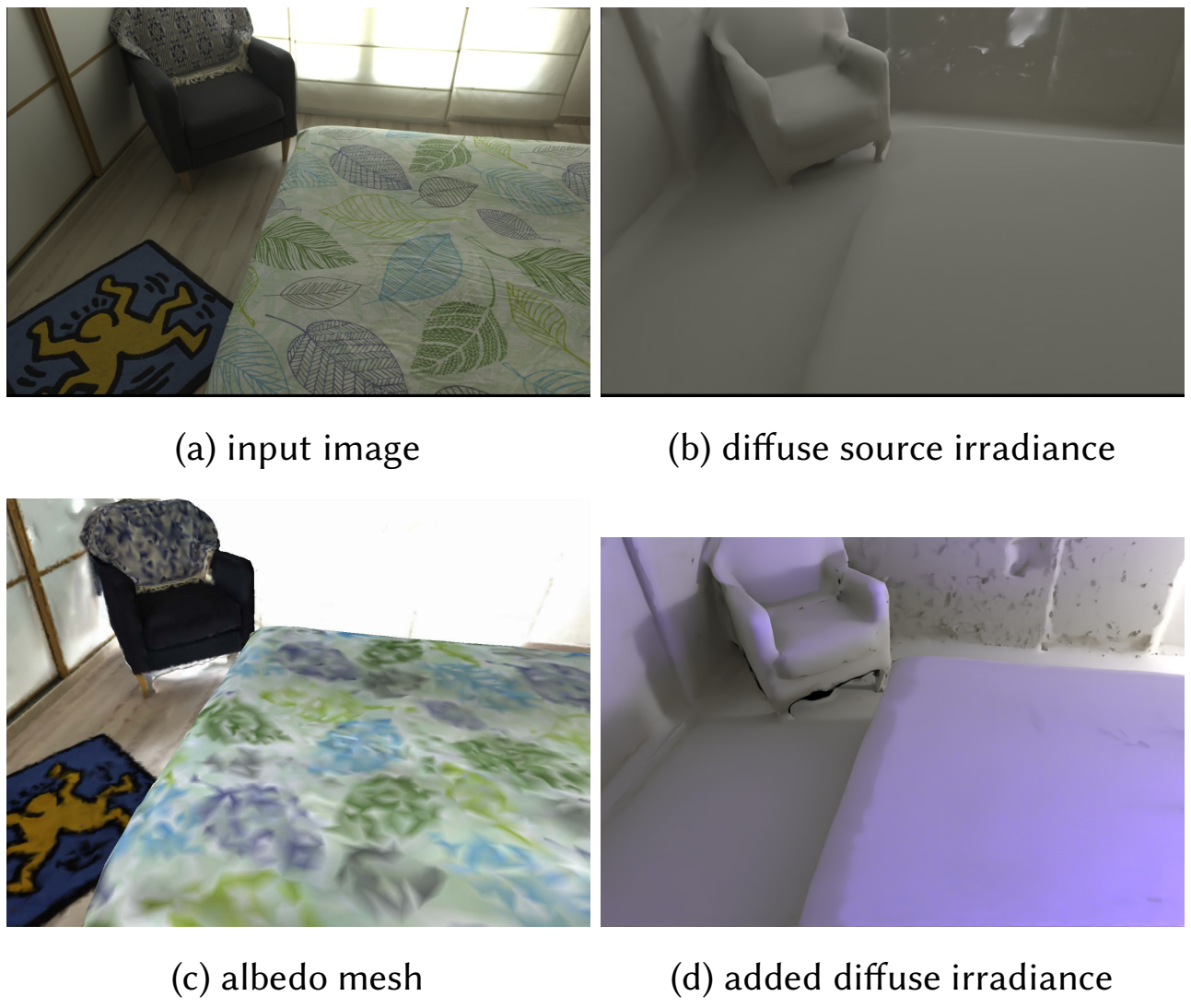

Fig. 4. We estimate the diffuse irradiance (b) corresponding to the input illumination by stochastic integration, sampling the input views. The illumination is fairly neutral in this scene, but our irradiance maps record full RGB information. This allows us to compute an approximate albedo by dividing the input image data (a) by the irradiance, which we store at the vertices of the 3D mesh (c). When adding a new light in the scene, we can compute the additional diffuse irradiance by path-tracing using the albedo mesh (d). In this example, we added two lights: a purple light to the right and below the camera, and a white light above and to the right.

*4.1.2 Diffuse albedo mesh.* Using our estimate of the diffuse source irradiance, we compute an albedo map $I_i/E_i^{src}$ for each view. We project all the albedo maps onto the MVS mesh and collect weighted averages of the reprojections that pass the depth test, into the mesh vertices. The weights are proportional to the inverse distance from the corresponding camera (see Fig. 4c for a visualization). This albedo is coarse but it is only an intermediate value to compute the diffuse irradiance from user-defined lights. In particular, our approach does *not* use any explicit material model; we let the network learn an implicit material representation.

*4.1.3 Path-traced added irradiance - $E_{1..n}^{add}$.* The user adds new lights by specifying area lights with constant emittance in 3D. For each input view we precompute a map of the irradiance from the light $E_i^{add}$ using bi-directional path tracing with the albedo mesh in Mitsuba [Jakob 2010]. Fig. 4d shows a target added irradiance map. For multiple lights, we exploit the linearity of light transport. We precompute irradiance maps $E_{1..n}^{add}$ for each light and linearly mix them at render time. This gives the user real-time control over the light intensities. For notational simplicity, the rest of the paper assumes a single light is added.

*4.1.4 Removed irradiance - $E_{1..n}^{rem}$.* The user also has control over the original scene lights in aggregate, using a single global intensity switch $\alpha_{\text{dim}} \in \{0, 1\}$. The amount of light to be removed is encoded relative to the source irradiance:

$$E_i^{rem} = \alpha_{\text{dim}} E_i^{src}. \tag{4}$$

When $\alpha_{\text{dim}} = 1$, the network completely removes the original illumination. When $\alpha_{\text{dim}} = 0$, the source lighting is fully preserved.

## 4.2 Mirror images for glossy reflections

We cannot expect a convolutional network — which is inherently local — to understand object materials and synthesize realistic reflections by correlating distant parts of the scene. So, we simplify the learning problem by precomputing first-bounce mirror images (§ 4.2.1) for each input viewpoint. Observing two versions of the scene — directly and with the pure mirror hypothesis — the network can estimate the glossy illumination and decouple it from the diffuse term (under our near-mirror assumption). We also compute a single target mirror image in the novel view (§ 4.2.2) that, unlike the source mirrors, accounts for the user-specified illumination changes. The target mirror helps synthesize glossy reflections in the new view.

*4.2.1 Source mirror images - $M_{1..n}^{src}$.* For each input camera $i$, we compute a source mirror image $M_i^{src}$ by casting a single ray per pixel into the scene. We trace the mirror ray at the first surface interaction, and record its intersection with the mesh. Then, we re-project this intersection (if any) into all the other views and retain the re-projection whose viewing direction is closest to the reflected ray direction. We sample this view's color at the reprojected intersection. We make sure to exclude viewpoints for which the intersection is not visible using a depth test. See Figure 5 for an example source mirror image and Figure 6, for an illustration of the procedure.

*4.2.2 Target mirror image - $\mathbf{M}^{tgt}$.* The target mirror image is computed dynamically and only for the novel view. Unlike the source mirrors, it needs to account for the user-specified lighting modifications. The sampling procedure is similar to § 4.2.1, except this time the target mirror image is sampled from:

$$\frac{I_i}{E_i^{src}}\left(E_i^{add} + E_i^{src} - E_i^{rem}\right), \tag{5}$$

where $\frac{I_i}{E_i^{src}}$ is a pseudo-albedo term, and $\left(E_i^{add} + E_i^{src} - E_i^{rem}\right)$ the modified irradiance. This ray-tracing step runs in real time using Optix [Parker et al. 2010].

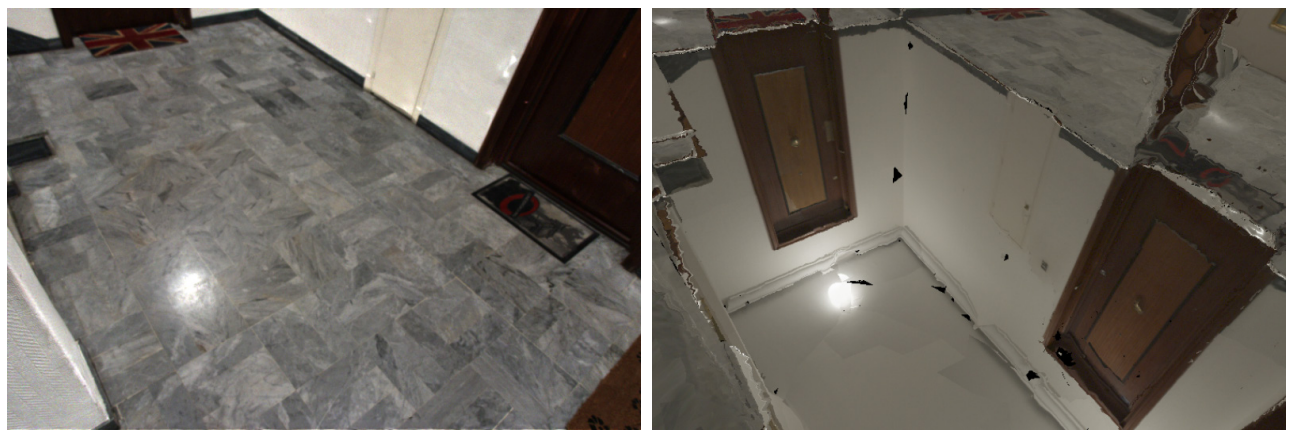

(a) input view (b) source mirror image

Fig. 5. For in each input view (a), we compute a mirror image (b) by tracing a ray in the mirror direction of the first mesh intersection and sampling the image whose camera—intersection vector makes the smallest angle with the mirror direction (see also Figure 6). Mirror images help the network analyze scene materials, and identify complex glossy effects like the salient highlight on the floor.

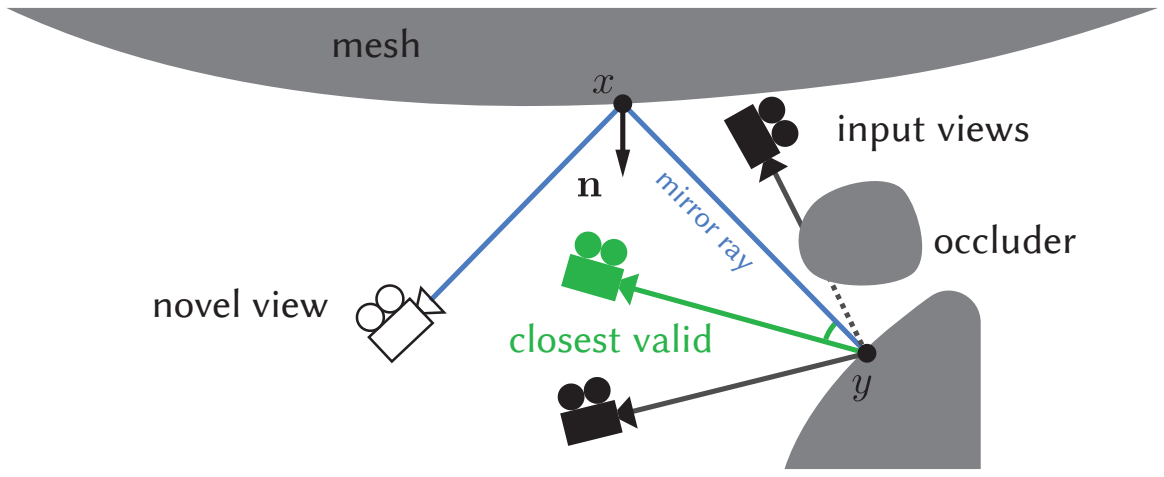

Fig. 6. To compute the mirror value at a reprojected point $x$ in a given view, we sample pixels from the input view that best align with the mirror direction, making sure to ignore cameras that do not see the intersection $y$ between the mirror ray and the geometry using a visibility test.

## 4.3 Interactive reprojection in the novel view

The precomputed feature maps described so far are stored in the reference frames of the $n$ input cameras. To synthesize a novel view, we need to reproject them into the common coordinate system of the new view. Reprojecting and using *all* the input views would be cumbersome because it would lead to large and variable number of network inputs at run-time. Besides, some views can be redundant or uninformative, e.g., they may not overlap at all with the novel view. Instead, we assemble a fixed number of composites by reprojecting the feature maps on-the-fly at run-time and blending them using weighted averages. We denote these composite reprojections with bold font variables.

We reproject the view-dependent features $I_{1..n}$ and $M_{1..n}^{src}$ and blend to produce 8 composites each: $\mathbf{I}_{1..8}$ and $\mathbf{M}_{1..8}^{src}$ respectively. For this we propose two reprojection algorithms. The first (§ 4.3.1) tries to maximally exploit the available texture resolution. It uses four well-known Image-Based Rendering heuristic blending weights to produce the first 4 composites. Second, the remaining 4 composites seek to maximize color entropy, i.e., to collect varied observations of the view-dependent effects. They are obtained using rank filters on the luminance of source images (§ 4.3.2).

Irradiance features $E_{1..n}^{src}$, $E_{1..n}^{add}$ and $E_{1..n}^{rem}$ do not change with the viewpoint. We reproject them using a simpler third algorithm (§ 4.3.3) to produce $\mathbf{E}^{src}$, $\mathbf{E}^{add}$, and $\mathbf{E}^{rem}$ respectively.

*4.3.1 Reprojections maximizing texture quality.* For the first group of 4 composites, we try to maximize the reprojected image resolution by sampling pixels from viewpoints that are close to the novel view. We also try to limit the number of pixels with no data, and to avoid temporal discontinuities (popping artifacts) when changing viewpoint. To achieve these goals, we use 4 distinct blending weights $w_i$ inspired by Buehler et al. [2001], for $i \in 1, \ldots, 4$ to assemble the 4 composites image $\mathbf{I}_{1..4}$, and similarly for the source mirror images $\mathbf{M}_{1..4}^{src}$.

Let $c_{\text{new}}$ be the 3D position of the novel camera, and $c_i$ that of the camera $i$ to be weighted. Consider a pixel in the novel view, and let $x$ denote the 3D position of the corresponding scene point, with normal $n$. Our four weights are given by:

$$w_1 = 1/\|c_{\text{new}} - c_i\|_2^2, \quad w_2 = \left(\frac{c_{\text{new}} - x}{\|c_{\text{new}} - x\|_2} \cdot \frac{c_i - x}{\|c_i - x\|_2}\right)^2,$$

$$w_3 = 1/\|c_i - x\|_2^2, \qquad w_4 = \exp((c_i - x) \cdot n/(0.1\|c_i - x\|_2)) - 1.$$

Only $w_1$ and $w_2$ are view dependent (i.e., depend on $c_{\text{new}}$), but they smoothly blend all the inputs which avoids popping. $w_1$ favors cameras close to the novel view. $w_2$ favors cameras that see the point from a direction similar to the novel view. $w_3$ favors cameras close to the 3D point, which maximizes texture resolution. Finally $w_4$ favors cameras that observe the surface at $x$ orthogonally, which reduces projection errors due to erroneous depth. These weighting schemes also provide observations of different material behavior: for instance $w_4$ selects views with less Fresnel effect. The top row of Figure 7 illustrates these reprojections for the wooden part of the bed.

*4.3.2 Reprojections optimized for material characterization.* The second group of 4 composites maximizes viewpoint diversity. They provide varied observations of the non-diffuse scene surfaces from cameras that are not clumped together. For each pixel $p$ in the novel view, we reproject all the inputs $I_{i..n}$ and sort them by luminance value. The composites are constructed by stitching the reprojected pixel colors, with the following selection criteria: $\mathbf{I}_5$ picks its color from the view with the highest luminance, $\mathbf{I}_6$ from the second highest luminance, $\mathbf{I}_7$ from the median luminance and $\mathbf{I}_8$ from the lowest luminance. This allows the network to see large variations in luminance, increasing the probability of observing non-diffuse material appearance.

Intuitively, the minimum luminance is often very close to the diffuse component of the surface [Szeliski et al. 2000]. On the contrary, the two highest luminance provide observations of bright specular highlight, if any. The discrepancies or agreement between these two extremes gives insight on a material's glossiness. Finally, the median is robust to outliers. When the geometry is poorly reconstructed or missing, Image-Based Rendering methods still sample pixels corresponding to missing geometry. This leads to "floaters": artifacts that appear to float in the air. The median filter discards floaters. The bottom row of Figure 7 shows the composite maps of this second group.

We obtain the source mirror composites $\mathbf{M}^{src}_{5..8}$ similarly: the luminance of $I_{i..n}$ is used for ranking, like before, but the values are sampled from $M^{src}_{1..n}$ this time.

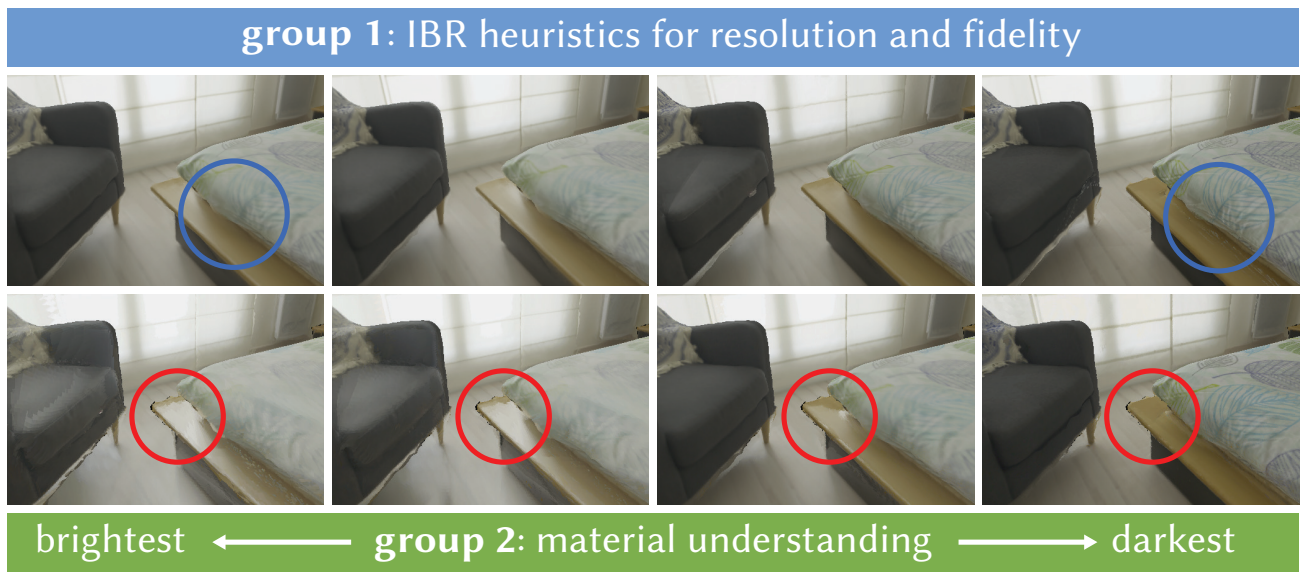

Fig. 7. We assemble a total of 8 reprojections from the input images, divided in two groups. The first group (top) tries to maximize reprojected resolution, image fidelity and coverage. For instance favoring cameras close to the novel view reduces reprojection errors but favoring cameras close to the observed point provides higher texture quality (blue circles, top). The second group (bottom) gather pixels that span the largest possible range of luminance values to help material understanding. In particular, these reprojections capture brightness variations due to specular highlights (red circles, bottom).

*4.3.3 Reprojecting irradiance maps.* Irradiance maps do not depend on the viewpoint so we use a simpler blending procedure to reproject them in the new view. The composite is a weighted average of all the $E^{src}_i$ (resp. $E^{add}_i$, $E^{rem}_i$) that pass the depth test, reprojected in the new view, with weight inversely proportional to the distance between camera $i$ and the novel view camera. This procedure yields three composite reprojections: $\mathbf{E}^{src}$, $\mathbf{E}^{add}$, and $\mathbf{E}^{rem}$.

### 4.4 Implementation details

We apply minimal post-processing to the MVS output and compute additional auxiliary network inputs.

*Geometry processing.* We simplify the mesh and filter normals using Laplacian smoothing. We identify planar surfaces using RANSAC on the mesh vertices snapping normals below a noise threshold to the plane. This improves overall robustness of our approach. Further post-processing could improve the rendering quality but this is beyond the scope of our method.

*Additional features.* For each pixel in the novel view we compute the normalized inverse depth (disparity) and the surface normal at the corresponding mesh point. This provides some geometric understanding. We also store the angle between surface normals and the view vector to help the network learn the Fresnel reflection effect. Finally, we compute the ratio of distances between: (1) the observed and the reflected points, and (2) the observed points and the camera projection center. This helps the network smooth reflections on rough surfaces. We denote these additional feature maps, collectively, by $\mathbf{F}_{\text{extra}}$.

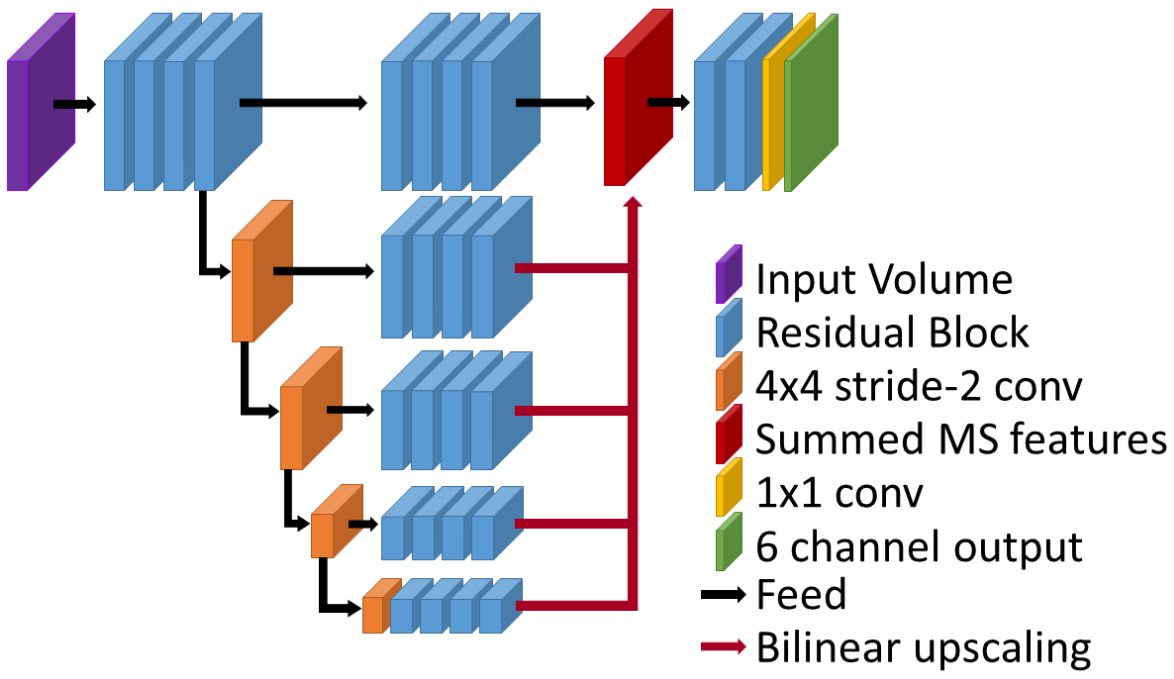

Fig. 8. Our network uses a multi-scale architecture to process the input features maps. It produces two output images, for the diffuse and specular image component respectively.

## 5 NETWORK AND TRAINING PROCEDURE

The network input is the concatenation of the composite reprojections, target mirror image and auxiliary features described in Section 4, namely $[\mathbf{I}_{1..8}, \mathbf{M}^{src}_{1..8}, \mathbf{M}^{tgt}, \mathbf{E}^{src}, \mathbf{E}^{add}, \mathbf{E}^{rem}, \mathbf{F}_{\text{extra}}]$. It produces two output images $O_{\text{diffuse}}$ and $O_{\text{vdep}}$ for the diffuse and specular image content, respectively.

### 5.1 Network architecture

Our network uses a multi-scale architecture shown in Figure 8. A 4-block ResNet [He et al. 2016] processes the input. Its output is then decomposed into a 5-level feature pyramid, using convolution layers with kernel size 4 and stride 2 as a downsampling operator. Each pyramid level is processed by an independent 4-block ResNet. The outputs of the 5 pyramid levels are bilinearly upsampled back to the original resolution and summed. A final 2-block ResNet, operating at full resolution, processes the summed features to produce two outputs: $O_{\text{diffuse}}$ containing the diffuse image component, and $O_{\text{vdep}}$ containing the view-dependent specular effects.

We use fixup ResNets [Zhang et al. 2019], with 6 residual blocks and 66 filters per convolutional layer, with kernel size 3, zero padding and ReLU activations. The input feature maps make up 66 channels as well. We found the “fixup” initialization to significantly improve convergence speed and output quality. The diffuse and specular components are supervised separately (§ 5.3), then summed to obtain the final relit image (§ 5.2).

### 5.2 Input and output tonemapping

We found training in a tonemapped rather than linear space to be more stable, as observed previously [Gharbi et al. 2019]. Tonemapping ensures the distribution of radiance values is not too skewed toward dark pixels. We tonemap all the radiance inputs to the network ($\mathbf{I}_{1..8}$, $\mathbf{M}^{src}_{1..n}$ and $\mathbf{M}^{tgt}$) using the operator proposed by Kalantari

and Ramamoorthi [2017]:

$$x \mapsto \frac{\log(1+\mu x)}{\log(1+\mu)}, \text{ with } \mu = 64. \tag{6}$$

The network's outputs are in tonemapped space as well. The target for the diffuse term is the tonemapped ground-truth radiance $O^\star_{\text{diffuse}}$, while our specular output is trained to estimate the following residual term, in tonemapped space:

$$O^\star_{\text{vdep}} := O^\star - O^\star_{\text{diffuse}}, \tag{7}$$

where $O^\star$ is the ground-truth total radiance (i.e., diffuse and specular). We obtain our final tonemapped rendering by summing the two terms: $O = O_{\text{diffuse}} + O_{\text{vdep}}$.

### 5.3 Loss functions

We train the network to minimize the sum of several losses:

$$\mathcal{L}_{\text{total}} = \mathcal{L}_1(O_{\text{diffuse}}, O^\star_{\text{diffuse}}) + \lambda_{\text{vgg}}\mathcal{L}_{\text{vgg}}(O_{\text{diffuse}}, O^\star_{\text{diffuse}}) \tag{8}$$
$$+ \lambda_{\text{vdep}}\mathcal{L}_1(O_{\text{vdep}}, O^\star_{\text{vdep}}) + \lambda_{\text{vgg}}\mathcal{L}_{\text{vgg}}(O_{\text{vdep}}, O^\star_{\text{vdep}}) \tag{9}$$
$$+ \mathcal{L}_1(O, O^\star) + \lambda_{\text{vgg}}\mathcal{L}_{\text{vgg}}(O, O^\star) \tag{10}$$
$$+ \mathcal{L}_{\text{multiview}}(O_{\text{diffuse}}) \tag{11}$$
$$+ \lambda_{\text{adv}}\mathcal{L}_{\text{adversarial}}\Big([O_{\text{diffuse}}, O_{\text{vdep}}], [O^\star_{\text{diffuse}}, O^\star_{\text{vdep}}]\Big), \tag{12}$$

where $\star$ denotes the synthetically rendered ground truth targets.

Equations (8), (9) and (10) are data fidelity terms on diffuse, view-dependant and summed signals respectively. $\mathcal{L}_1$ is a standard $L_1$ loss and $\mathcal{L}_{\text{vgg}}$ is a perceptual loss based on the $L_1$ distance between VGG19 activations [Johnson et al. 2016]. We use the activations of the first five maxpooling operators, weighted by 0.5, 0.25, 0.05, 0.05 and 4.0 respectively. The last weight is higher to put more emphasis on deeper features, because they are resilient to misalignment errors arising from the reprojection [Wei et al. 2019]. We found this weighting scheme to be crucial for sharp outputs, recovering sharper outputs than using $\mathcal{L}_1$ loss alone and speeds up the network convergence. We set $\lambda_{\text{vgg}} = 0.1$. The specular term is overall much darker than the diffuse component, so we set $\lambda_{\text{vdep}} = 10$ to balance its contribution to the loss.

$\mathcal{L}_{\text{multiview}}$ is a multi-view consistency loss on the diffuse component implemented as the $L1$ difference between the reprojections of nearby views onto one another. These reprojections use the ground truth optical flow produced by the renderer. This loss penalizes residual view dependent effects that may otherwise creep in and would make our output temporally unstable.

$\mathcal{L}_{\text{adversarial}}$ is an adversarial loss on the concatenated network outputs that improves overall texture sharpness and the realism of our outputs. Our discriminator is a global relativistic GAN [Jolicoeur-Martineau 2019] with the architecture of Radford et al. [2015]. We use $\lambda_{\text{adv}} = 0.05$.

### 5.4 Optimization details

We train the network and discriminator using RMSProp [Hinton et al. 2012] in PyTorch with default parameters. We use a learning rate of $10^{-4}$ with a batch size of 1 for 600k iterations, which takes approximately 96 hours on a NVIDIA RTX 6000 GPU.

## 6 SYNTHETIC TRAINING DATA

Acquiring real-world data to supervise a relighting model is difficult. Like previous work [Philip et al. 2019], we train our model on computer-generated imagery. For the model to generalize to real-world captures, we need realistic and varied geometry, materials and lighting, so we start from artist-modeled 3D scenes, either purchased [1] or available freely [2]. We convert the scenes to a common format and use Mitsuba [Jakob 2010] for rendering. We used a total of 16 synthetic scenes for training.

*Lighting and cameras.* We manually place cameras and lights in each scene using an interactive interface and render an image for each viewpoint and each lighting condition. In total, our training set contains around 2000 individual viewpoints, with 6–7 distinct lighting conditions per scene (including the scene's original lighting). We normalize the intensity of all our lights to 1.0. At training time, we linearly combine multiple lighting conditions, randomizing the mixing weights, for data augmentation.

*Rendering the target images.* We render two images for each training viewpoint/illumination, corresponding to the diffuse component $O_{\text{diffuse}}$ and glossy term $O_{\text{vdep}}$, respectively (§ 5). These are rendered using bi-directional path-tracing with 256 samples per pixel, and denoised using Optix. Figure 10 shows an example pair.

*Simulating realistic input data.* We use the pipeline described in Section 4 to create the input training data. Synthetic training scenes have exact geometry which, if used directly, would lead to a domain gap at test time because of inaccuracies in the MVS reconstruction of real scenes. To avoid this, we render additional images and use them to reconstruct an MVS proxy for each synthetic scene [Philip et al. 2019]. We use the inaccurate meshes (see Fig. 9) to compute our input training data. Because the target images are rendered with the true mesh, our trained network is robust to small errors in the geometry.

*Altering synthetic scenes for MVS.* We found that MVS reconstruction sometimes fails on synthetic indoor renderings because of large textureless surfaces (walls, ceilings) or purely specular materials with no real-world imperfections (e.g., mirrors). So, for the MVS reconstruction step only, we replace scene materials with unrealistic but detailed diffuse textures. We show a training scene modified for MVS reconstruction in Fig. 10c.

## 7 EVALUATION

We captured multi-view datasets for qualitative analysis on real-world scenes (§ 7.1). To the best of our knowledge, no previous method can jointly relight scenes and enable free-viewpoint rendering for indoor scenes with glossy materials. We thus compare to view-synthesis (§ 7.3) and relighting (§ 7.2) techniques separately. We run our quantitative comparisons and model ablations (§ 7.4) on two held-out synthetic scenes, shown in Figure 15 and in supplemental. Figure 1 and 11 shows our model's output for relighting and view-synthesis on five real scenes. Our results are best

[1] We use the scenes from evermotion.org Archinteriors 30: 001-010, Archinteriors 01: 004 and 005.

[2] We use the scenes *Country Kitchen* and the *White Room* from [Bitterli 2016].

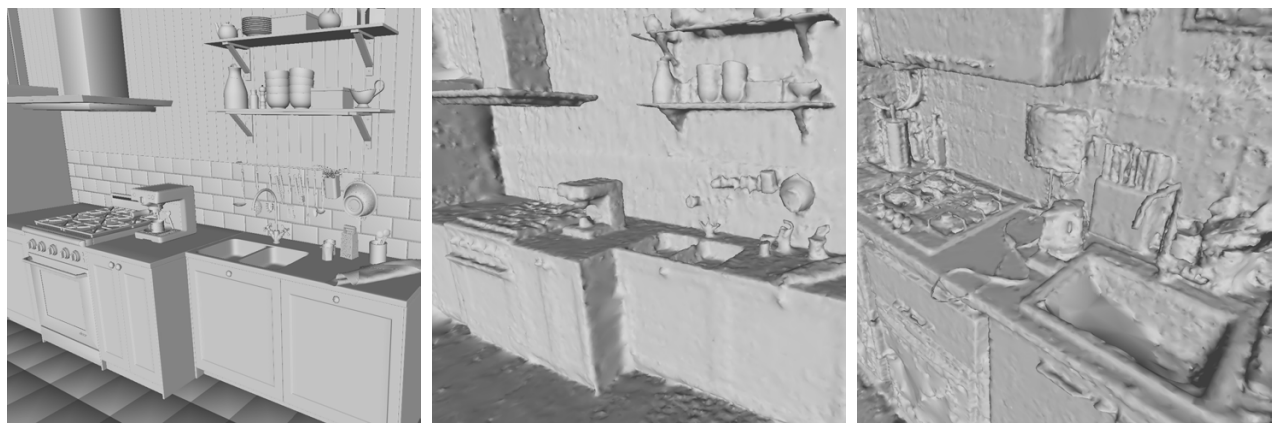

(a) ground truth mesh for a training scene (b) reconstruction of the same training scene (c) reconstruction of a real scene

Fig. 9. To minimize the domain gap between real data and our synthetic training scenes, we compute the training network inputs on a degraded mesh, obtained by multi-view stereo from the rendered images instead of the ground truth mesh (a). The training reconstructions (b) exhibit artifacts that are qualitatively very similar to what we expect to see in real scenes (c). This helps the network generalize well to real scenes.

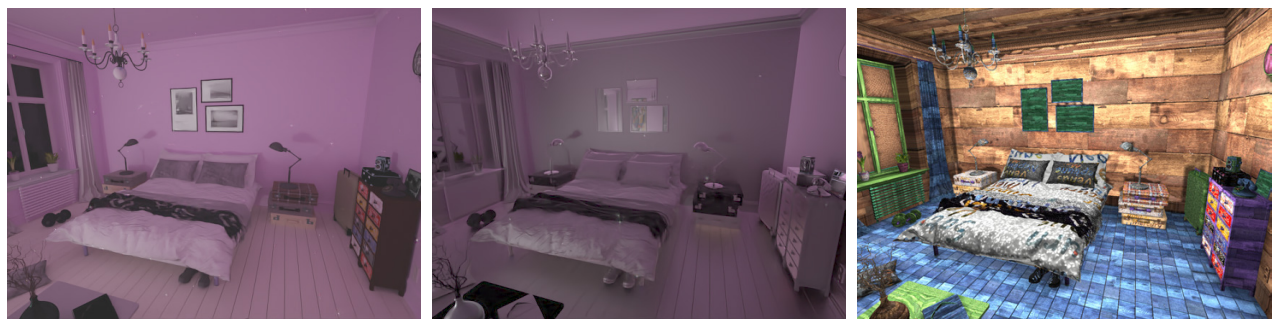

(a) synthetic rendering (diffuse) (b) glossy component (brightened for display) (c) automatic scene modification for MVS reconstruction

Fig. 10. We decompose our synthetic target renderings into diffuse (a) and glossy components (b). To better simulate real data, the network inputs used at training time are computed using an MVS reconstruction of the scene instead of the ground truth geometry. We facilitate the reconstruction using an automated tool that changes the lighting and adds markings and textures (c) (note this mesh used *only* for reconstruction).

appreciated in the supplemental video and website (https://repo-sam.inria.fr/fungraph/deep-indoor-relight/) , where all paths are provided. As can be seen in the figures and videos, our method allows plausible relighting of scenes with complex, glossy materials, while allowing free-viewpoint navigation. Our method provides faithful flow of highlights when moving the camera, both for the original (real) light sources, and those added for relighting.

### 7.1 Real-world scene capture

We created a dataset of real test scenes, using a Canon 50D DSLR and a 18-50mm lens, photographing in 14-bit RAW format from 250–350 viewpoints per scene. Specifically, we captured the following scenes, shown in Fig. 1, 12-11 and in the supplemental material: Bedroom1 (352 photos), Bedroom2 (306), LivingRoom (321), Sofa (282), Kitchen (274), Hall (291). All these scenes were reconstructed using Structure from Motion and MVS using the commercial RealityCapture [RealityCapture 2019] software, providing calibrated cameras for each input photo and an approximate 3D geometric proxy mesh.

Our irradiance estimation requires that the real light sources be directly visible in at least some of the photos. The linearity and higher dynamic range of the RAW format helps us estimate relative light powers. Despite the higher bit-depth, directly visible light sources can still be clipped. In this case, the user needs to provide a few clicks to allow estimation of the clipped source irradiance; we describe this procedure in Appendix A. We manually place additional lights in the reconstructed 3D scene when relighting.

### 7.2 Relighting

To our knowledge, there is no previous method that can relight multi-view datasets of entire scenes. Instead in Figure 12 we compare to a baseline that uses the 3D information available to our method, and the state of the art inverse rendering technique for indoor scenes from Li et al. [2019]. Their method does not directly relight images; it produces an albedo from a single image. Our baseline assumes a Lambertian material model: it combines their albedo with our target irradiance (computed from the MVS mesh) to get the relit image. Note that we provide the benefit of irradiance computed using 3D information to the baseline, allowing a comparison that is as fair as possible. Compared to this baseline, our relighting solution preserves texture in the scene, and benefits from the overall quality of our neural rendering.

We also make a comparison to Wu and Saito [2017], which is indicative of modern single-image methods. We asked the authors to insert lights in the scene to produce an effect similar to our lighting modifications, Figure 13. Their approach only uses a single image and has a very approximate notion of 3D geometry compared to the much richer input we use. Compared to ours, their method cannot cast shadows for complex objects such as the armchair. It creates a static highlight for the novel light on the hardwood floor, without the view-dependent effects we can synthesize.

We also evaluate the robustness of our relighting to the quality of the reconstructed geometry. We reconstructed the scene in Fig. 11 (top rows) using images $\frac{1}{8}$ the resolution, which resulted in degraded geometry. We show the original and degraded geometry in Fig. 14(top) and the resulting relit images below. We can see that the quality of shadows is slightly inferior to that using the original geometry, but overall the relighting effects are well preserved.

### 7.3 View-synthesis

We compare our model to four free-viewpoint rendering methods: a baseline Unstructured Lumigraph Rendering (ULR) implementation with per-pixel blending weights [Bonopera et al. 2020; Buehler et al. 2001], Deep-Blending [Hedman et al. 2018], as well as the very recent Neural Radiance Fields — NeRF [Mildenhall et al. 2020] and Free View Synthesis (FVS) [Riegler and Koltun 2020]. We used the original Deep-Blending network and code [Bonopera et al. 2020] and the authors' implementation of NeRF. The FVS comparisons were run by the authors using the model they trained on the Tanks and Temples dataset [Knapitsch et al. 2017], which has very different scene content compared to our scenes.

Figures 21,22 show we can synthesize realistic highlights that move coherently when changing viewpoints while previous methods cannot (see also supplemental videos). Since ULR, Deep Blending and FVS blend multiple frames and have no or limited material understanding, they tend to create double ghost highlights that appear to "jump" from one frame to the next (see video), or even completely suppress glossy reflections. ULR and our method use the same MVS

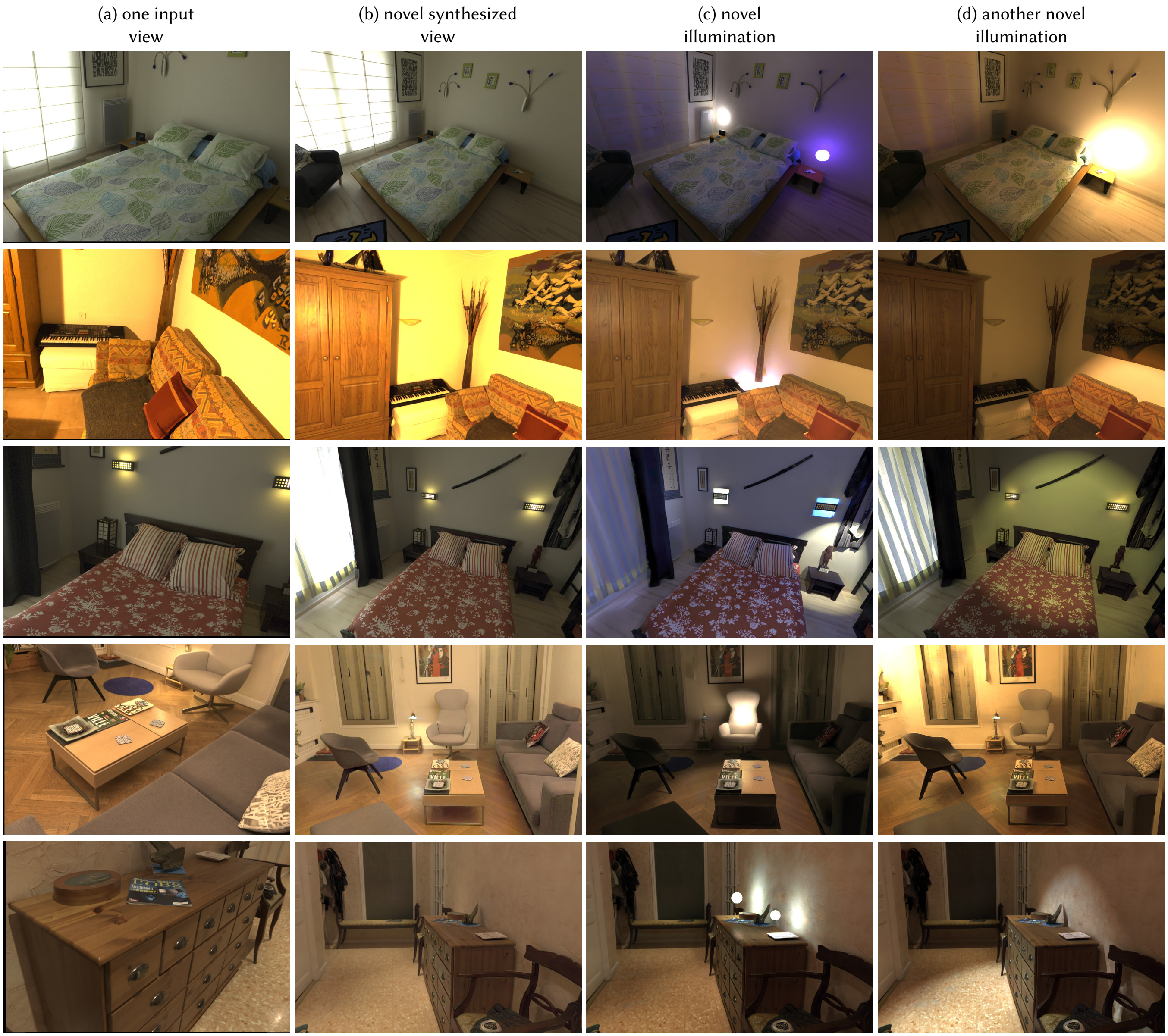

Fig. 11. A selection of view-synthesis and relighting outputs produced by our neural algorithm. For each row, we show one input view (a), a newly synthesized view (b) and two relightings, also at a novel viewpoint (c) and (d). The scenes are respectively: Bedroom2, Sofa, Bedroom1, LivingRoom and Hall.

mesh; this shows our neural rendering is largely robust to errors in the 3D reconstruction, as long as MVS can provide reasonable quality.

Hedman et al. [2018] use more complex *per-view meshes* that significantly improve the geometry, especially for small objects. As a result, Deep-Blending's rendering of small objects is sometimes more detailed than ours (e.g., the legs of the chair in Figure 22, first row). Still, our renderings are much sharper and realistic overall. Our method would benefit from per-view meshes or any other improvements to the MVS geometry, but this is largely orthogonal to our view-synthesis and relighting contributions.

As noted by Riegler and Koltun [2020], NeRF struggles with the unstructured capture required to reconstruct large scenes such as ours. For fair comparison, we made a best-effort attempt to improve NeRF's output by manually selecting a subset of images that produces the best visual and numerical quality for NeRF. We found much fewer, more tightly packed images, provide better visual quality *locally*, but they result in low coverage, and therefore worsen the *overall* image quality. NeRF can synthesize convincing moving highlights (see supplemental), but the overall image quality is low.

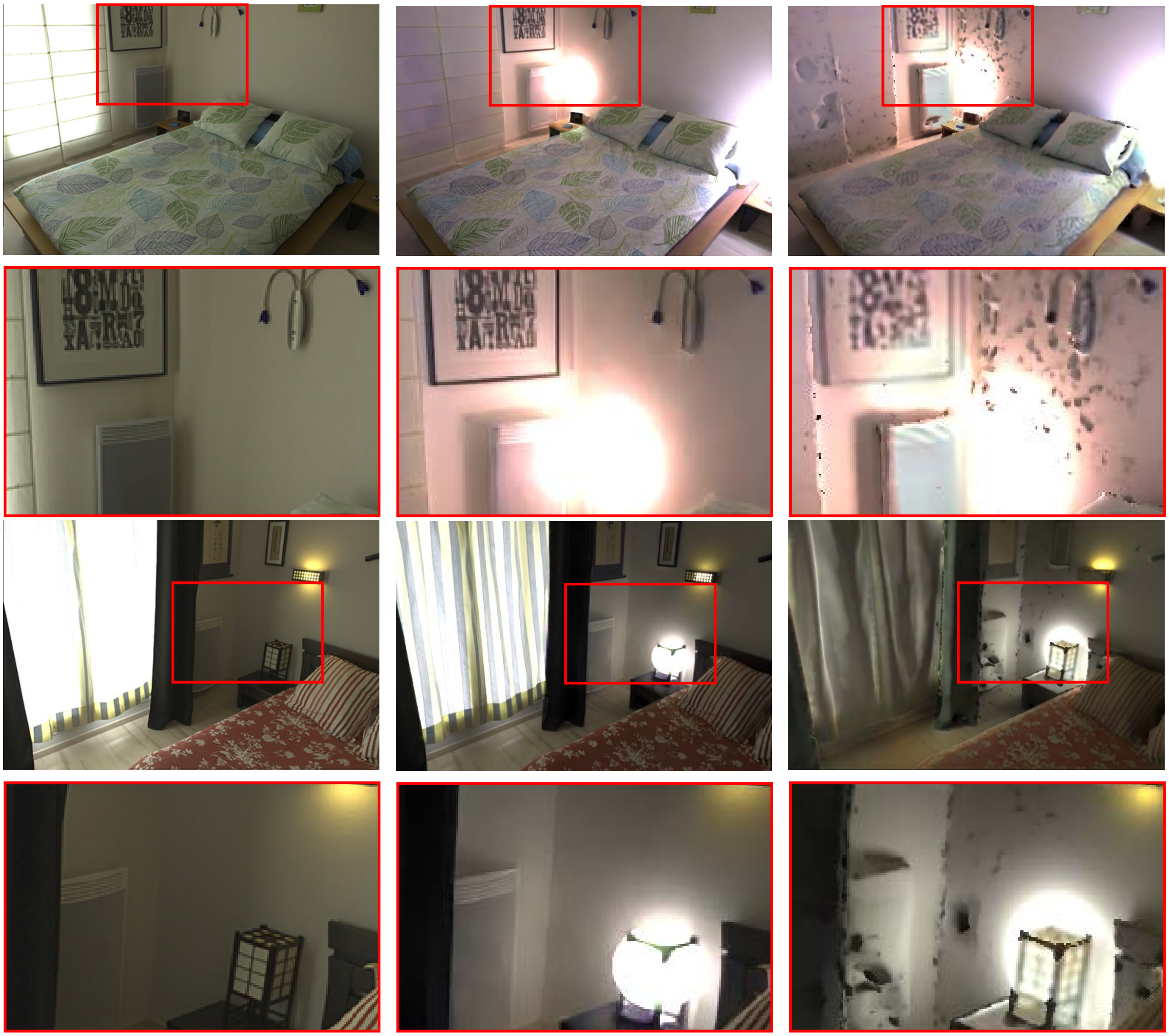

Fig. 12. **Relighting comparison** to a baseline using the method of Li et al. [2019]. Left: Input view with original lighting. Middle: Our relighting result. Right: Baseline result. For the baseline computation, we first extract albedo using the approach of Li et al. [2019] and directly multiply it with our target added irradiance to form a relit image. As we can see in the zoomed region, contrary to the baseline, our method correctly modifies the lighting under the stool, removes the shadows initially present and does not exhibit geometric artifacts while preserving high texture quality.

FVS [Riegler and Koltun 2020] gives good static results, but as with all previous view-synthesis techniques, does not render moving highlights correctly. They have more visual artifacts and less temporal coherence (see supplemental videos). This is possibly accentuated by the difference in content compared to their training dataset.

Table 1 summarizes our numerical evaluation on a path with 420 views in a held-out synthetic scene; Figure 15 and supplemental show the renderings. Appendix B analyzes this error as a function of the distance to input views. For completeness, we also performed a quantitative leave-one out evaluation of view synthesis, by holding out each one of the input images and performing novel view synthesis for our algorithm and previous methods, see Table 2. Such quantitative comparisons are not always very informative, since the outcome can depend on the metric (see BedRoom1 or LivingRoom), and in one case ULR [Buehler et al. 2001] has the best DSSIM score, even though it has clear visual artifacts in free-viewpoint synthesis. Nonetheless, for 4 out of 5 scenes, our method has a better LPIPS score than previous work.

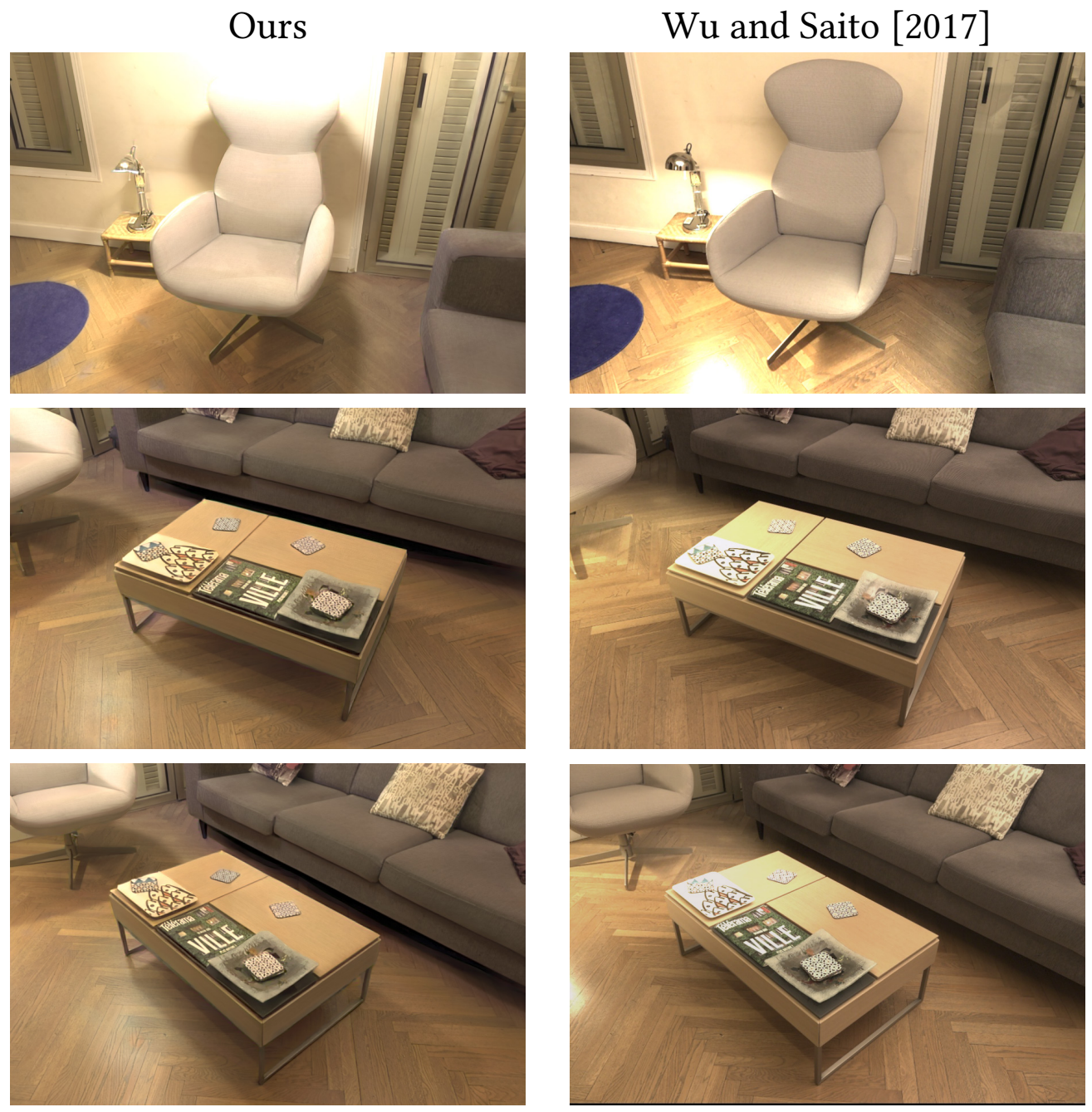

Fig. 13. **Single-image relighting comparison.** We provide an indicative comparison with the *single image* method Wu and Saito [2017] that has much less information that our multi-view input. We see that this method cannot reproduce shadows of complex objects such as the armchair, nor the view-dependent glossy effects on the hardwood floor.

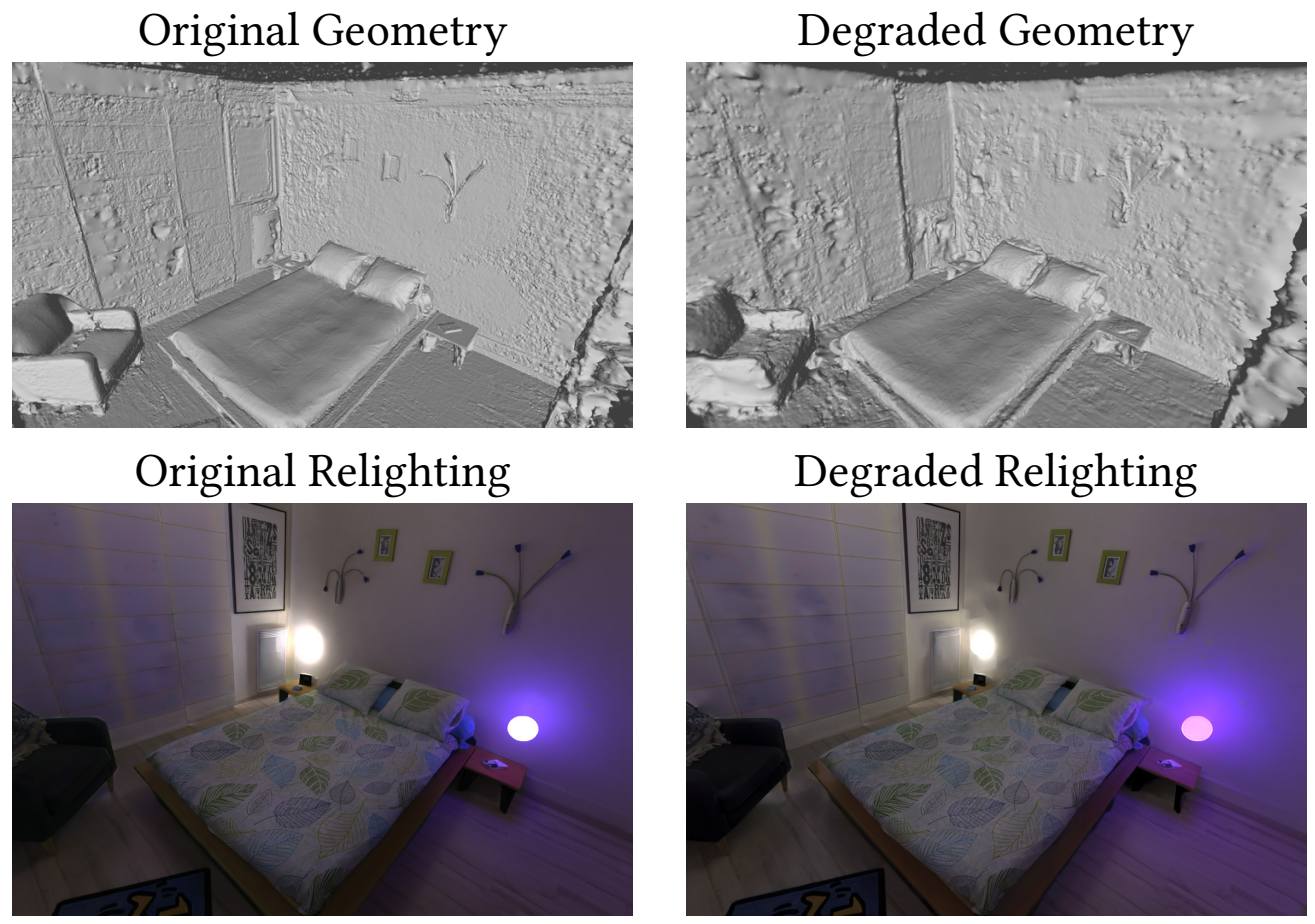

Fig. 14. **First row.** Left: original geometry; Right: geometry reconstructed with $\frac{1}{8}$ resolution, resulting in missing and incorrect reconstruction. **Second row.** Left: relighting with original geometry; Right: relighting with degraded geometry. Small errors can be observed in the shadows, e.g., from the armchair leg.

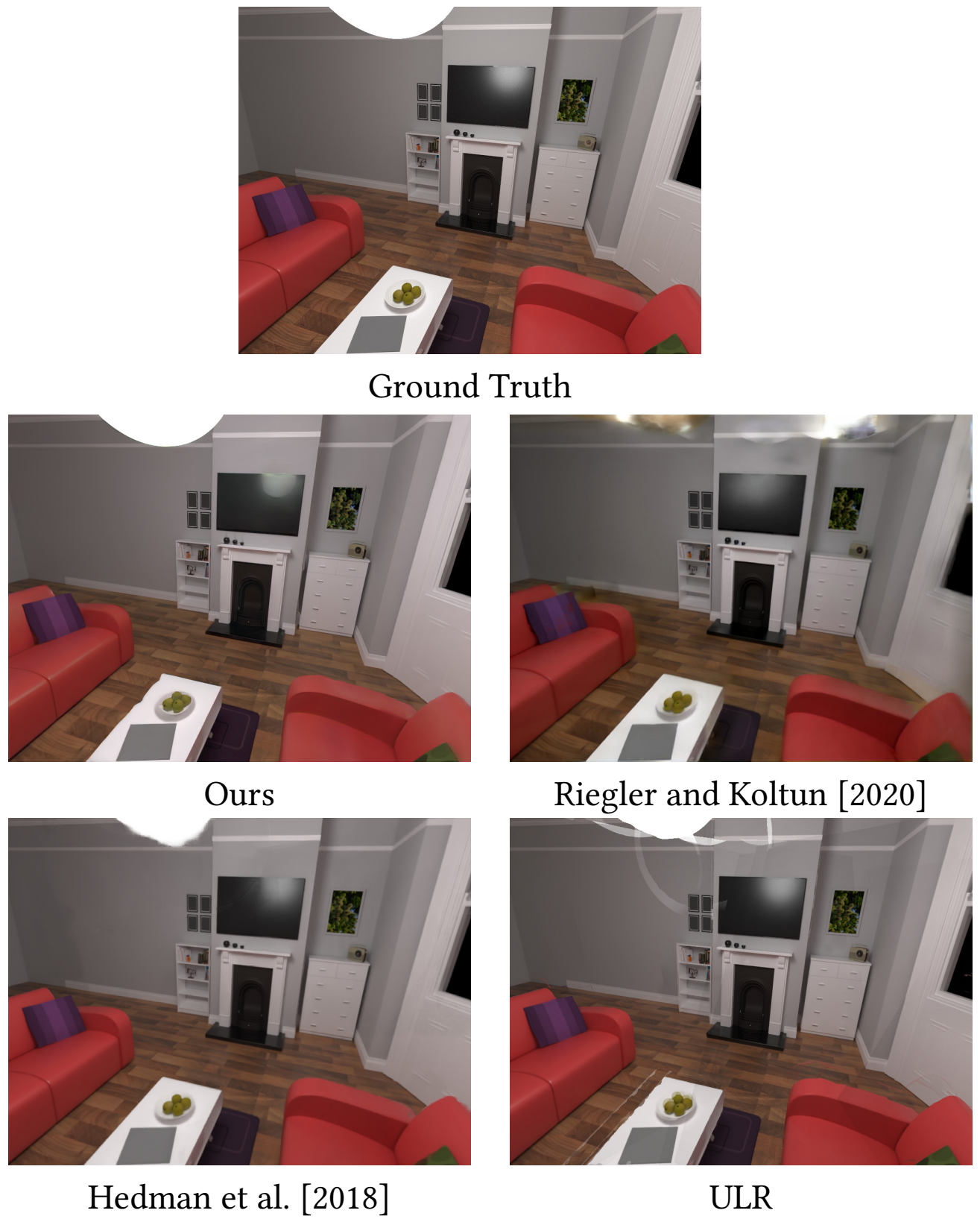

Fig. 15. **Held-out synthetic scene used for quantitative evaluation.** We show here the synthetic scene used for numerical evaluation. In rows 2 and 3, we show the synthesized result for each method; For all methods but ours, the highlight on the TV does not move correctly (see also supplemental).

### 7.4 Model ablations

We conduct an ablation study to isolate and highlight the importance of our input features and design choices. Every model variant in this section is trained from scratch using the same training procedure. Numerical results are summarized in Table 3. The values reported in Table 3 were computed over 1000 views along a path in each of the two scenes, with *different*, randomly selected input and output lighting conditions for each view. The table is sorted by the LPIPS metric. We see each component of our approach improves quality in different ways; e.g., both the GAN and use of VGG affect the perceptual LPIPS loss more than PSNR.

*7.4.1 Single reprojection.* Our reprojection scheme (§ 4.3) produces 8 composites that are key to learning a good implicit representation of scene materials. If we replace it with a single reprojection using $w_1$, the network is much less effective at separating specular and diffuse effects and copies incorrect glossy highlights baked in the reprojected input views, significantly reducing realism (Fig. 16).

Table 1. **View-synthesis evaluation.** Our method consistently outperforms previous work on a held-out synthetic scene.

| Scene / method | Synth Living Room | | | Synth Kitchen | | |
|---|---|---|---|---|---|---|
| | PSNR ↑ | DSSIM ↓ | LPIPS ↓ | PSNR ↑ | DSSIM ↓ | LPIPS ↓ |
| Ours | **29.25** | **0.042** | **0.112** | **31.85** | 0.036 | **0.152** |
| Hedman et al. [2018] | 27.07 | 0.045 | 0.148 | 28.19 | 0.039 | 0.184 |
| Buehler et al. [2001] | 24.50 | 0.052 | 0.142 | 29.04 | **0.032** | 0.155 |
| Riegler and Koltun [2020] | 21.66 | 0.073 | 0.208 | 27.17 | 0.062 | 0.263 |

Table 2. **Leave one Out View-synthesis evaluation on real test scenes.**

| Scene / method | Bedroom2 | | Sofa | | BedRoom1 | | LivingRoom | | Hall | |
|---|---|---|---|---|---|---|---|---|---|---|
| | LPIPS ↓ | DSSIM ↓ | LPIPS ↓ | DSSIM ↓ | LPIPS ↓ | DSSIM ↓ | LPIPS ↓ | DSSIM ↓ | LPIPS ↓ | DSSIM ↓ |
| Ours | **0.065** | **0.024** | 0.143 | 0.096 | **0.090** | 0.043 | **0.114** | 0.051 | **0.122** | **0.036** |
| Hedman et al. [2018] | 0.141 | 0.043 | **0.137** | **0.043** | NA | NA | 0.124 | **0.041** | 0.261 | 0.091 |
| Buehler et al. [2001] | 0.076 | 0.029 | 0.142 | 0.078 | 0.096 | **0.036** | 0.167 | 0.064 | 0.137 | 0.037 |
| Riegler and Koltun [2020] | 0.177 | 0.055 | 0.213 | 0.098 | NA | NA | 0.266 | 0.085 | 0.304 | 0.101 |

Table 3. **Ablations.** Quantitative comparisons on two held-out synthetic scene shows that each element in our design contributes to the final result quality. The table is ordered by increasing LPIPS.

| method | PSNR ↑ | DSSIM ↓ | LPIPS ↓ |
|---|---|---|---|
| ours | 24.46 | **0.051** | **0.120** |
| no split diffuse/Vdep | 24.95 | 0.052 | 0.129 |
| no target mirror | 23.85 | 0.056 | 0.135 |
| no source mirror | 23.78 | 0.057 | 0.135 |
| fewer training scenes | **24.96** | 0.056 | 0.138 |
| no GAN | 23.83 | 0.055 | 0.142 |
| no GAN nor VGG | 23.84 | 0.056 | 0.147 |
| no multiscale in network | 23.25 | 0.062 | 0.156 |
| no extra features | 22.56 | 0.060 | 0.158 |
| single reprojection | 22.72 | 0.065 | 0.159 |
| GT geometry for training | 23.25 | 0.065 | 0.173 |

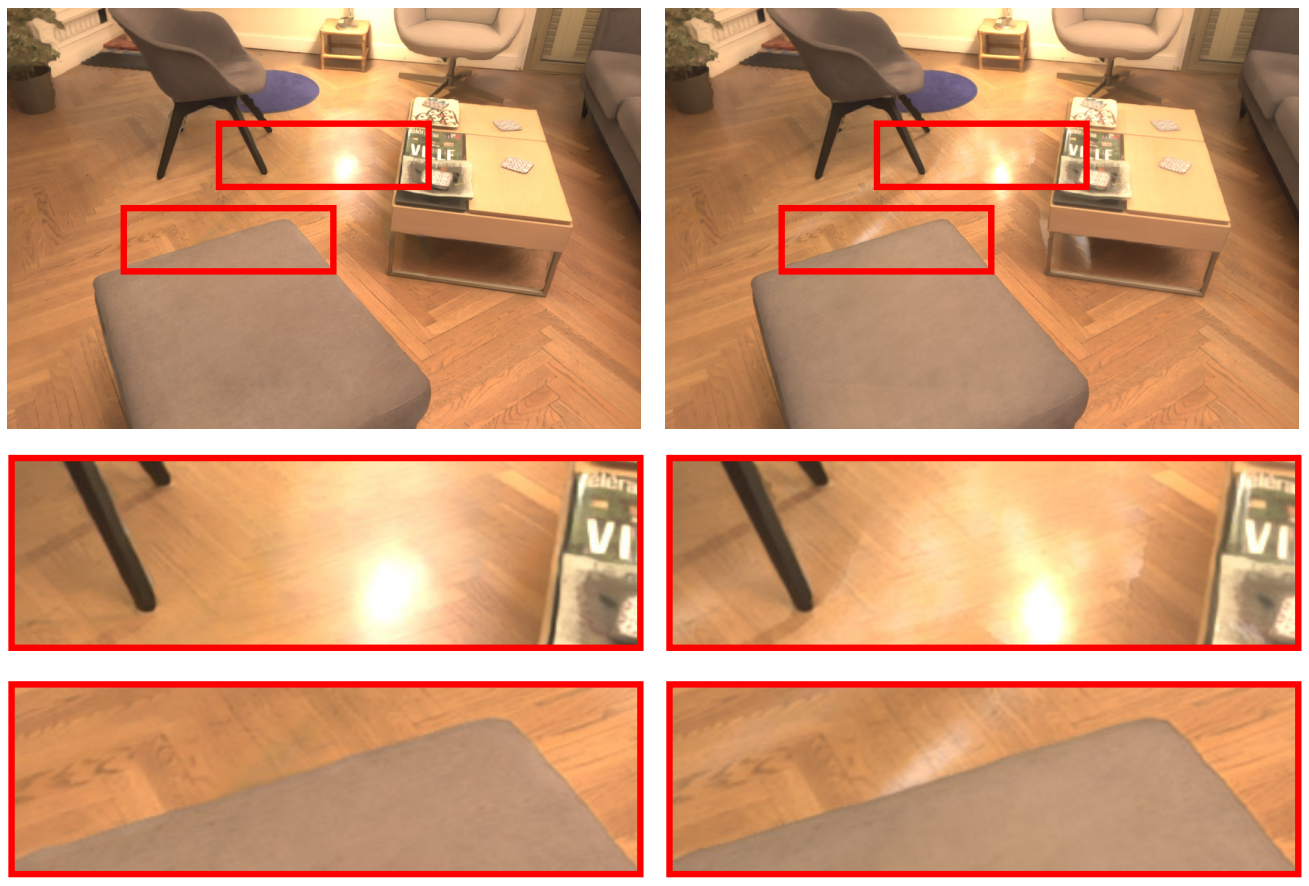

(a) our full solution (8 reprojections) (b) single reprojection

Fig. 16. **Single reprojection ablation.** With only one reprojection heuristic, using the closest views, the model cannot analyze scene materials properly (b) and outputs wrong colors corresponding to glossy highlights observed in the input views. In our full pipeline (a), the reflection is correct and there is no residual error from the reprojection.

*7.4.2 No target mirror image.* The target mirror image (§ 4.2.2) guides the synthesis of new glossy reflections. Without it, predicting accurate specular highlights, for both input and added light sources, is extremely difficult. This ambiguity leads to blurring (Figure 17).

*7.4.3 No separate supervision on diffuse/specular (*no split diffuse/Vdep *in table).* Supervising the diffuse and specular components separately (§ 5.3) leads to cleaner glossy reflections. If we supervise the summed output (Equation (10)) in an online manner, the network tends to focus on diffuse regions, which leads to unrealistic, muted reflections (Fig. 18).

*7.4.4 No Multiscale module in the network.* The Multi-scale module in our network is essential to obtain realistic specularities on rough surfaces. It yields a wider receptive field and can overcome the limitations of our mirror-like surface hypothesis, as shown in Fig. 19.

The remaining ablations in Table 3 show: the effect of removing the discriminator loss (*no GAN*), discriminator and VGG (*no GAN no VGG*), the source mirror layer and the use of extra features (§4.4). We also show the effect of only using ground truth (GT) geometry, i.e., no MVS reconstructions, and training using 8 instead of 16 scenes used for the full method.

## 7.5 Performance

We use RealityCapture [RealityCapture 2019] for the MVS reconstruction, which takes about 10 minutes per scene. Precomputing the source irradiance takes around 3 sec./image on average on an Intel Xeon E52650 @ 2.2Ghz, while the target added irradiance takes about 2 sec./image using Mitsuba and 16 samples per pixel on the same machine. Both images are denoised using Optix. So, for a

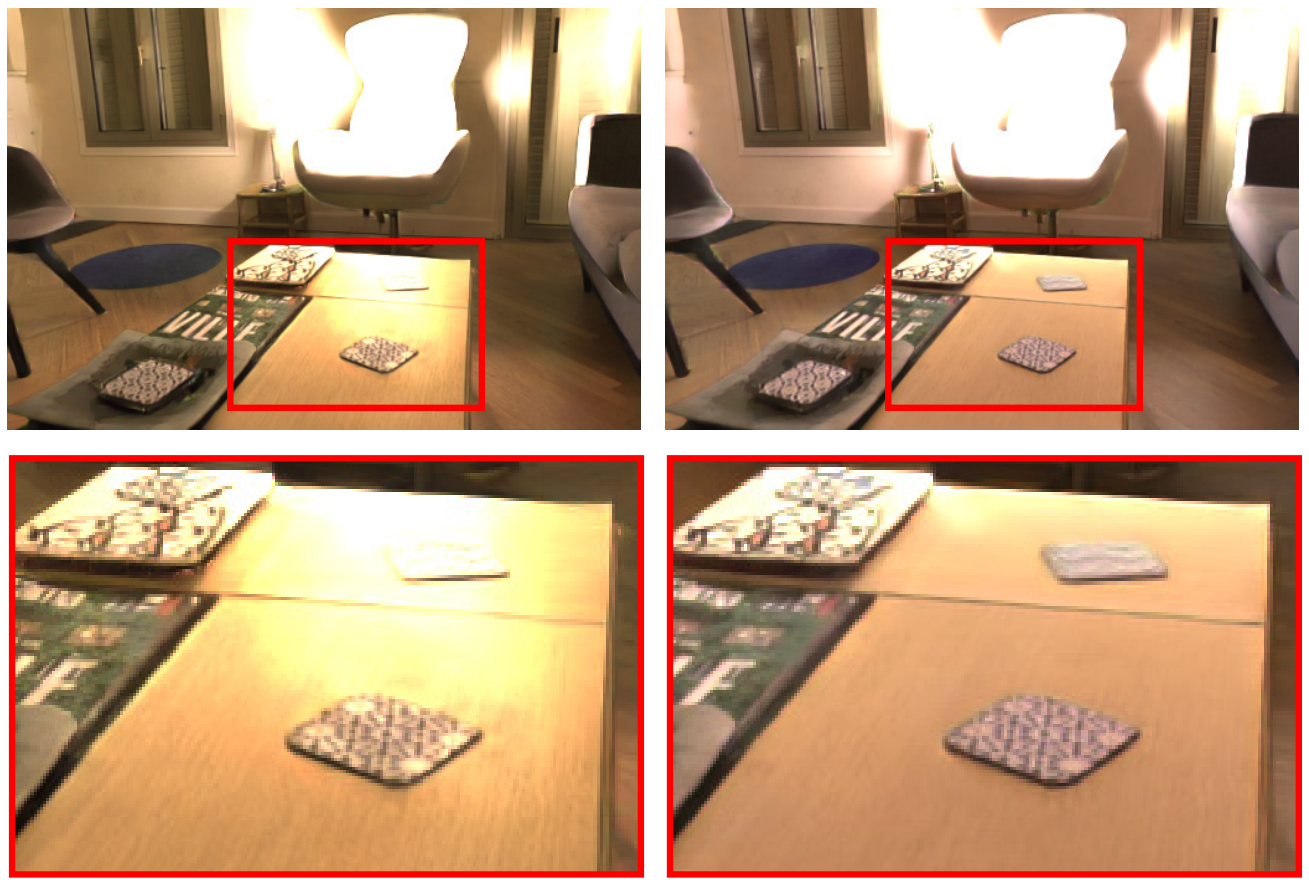

(a) our full solution (b) without target mirror

Fig. 17. **Target mirror ablation.** In our full pipeline (a), the target mirror image guides the reconstruction of realistic glossy reflections for existing and new light sources. Removing the target mirror (b) leads to lackluster images with no specular highlights.

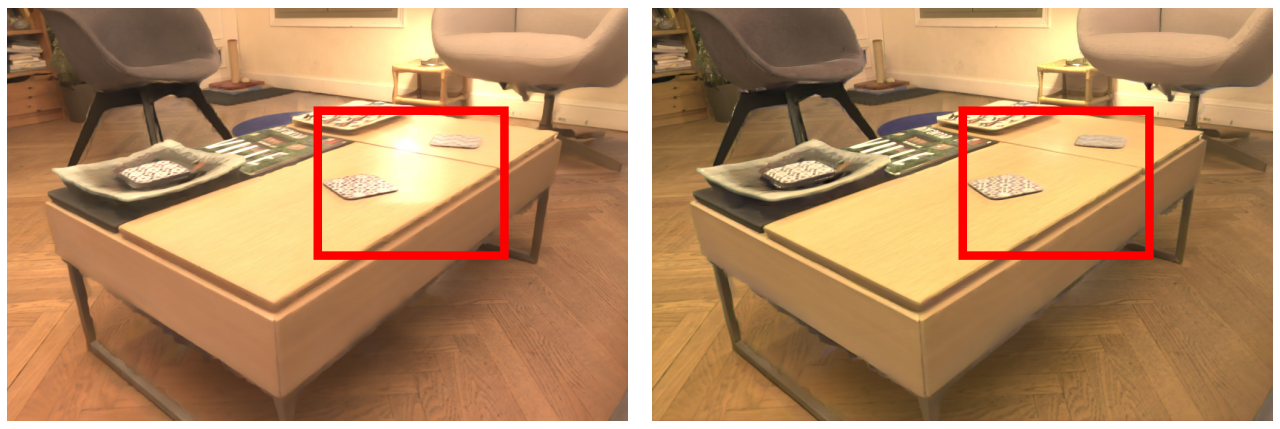

(a) our output, with diffuse/specular separation (b) without diffuse/specular separation

Fig. 18. **Ablation diffuse/specular supervision.** When separating the diffuse and specular output components, the loss can be better balanced, using scalar weights, to ensure correct highlights are synthesized (a). Without the split, the highlights are lost (b).

typical scene (3̃00 input frames), it takes about 10 minutes to precompute the added irradiance for a new lighting condition. Given this precomputation, our renderer is interactive and runs at 5–8 frames per second on a NVIDIA GeForce 2080RTX graphics card, rendering at $512 \times 384$ resolution. Our model can process images at arbitrary resolution, since it is fully convolutional. Using texture compression to lower memory costs, we can render up to $1024 \times 768$ images, maintaining the framerate above 2fps. The runtime cost is linear in the number of pixels. Note that most of the rendering time is taken by the network, whose inference could be optimized (e.g. 16 bit quantization, weight sparsification, etc).

## 8 LIMITATIONS

Our renderings sometimes exhibit residual artifacts, mostly from incorrect geometry due to failures in the MVS reconstruction. These are visible for example in Figure 15 on the coffee table or the leg of the chair in Figure 22, as well as in some video sequences (see supplemental), especially when light hits a surface at grazing angles, casting shadows from the incorrect geometry. MVS reconstruction may fail on strong highlights if too many pixels are clipped. In practice we found this was rarely an issue. Perfect mirrors and highly specular scenes are more challenging and would require dedicated capture setups [Whelan et al. 2018]. Both our reprojection and Monte Carlo integration can cause temporal flickering; this is largely mitigated by our multi-view consistency loss. Currently, we do not identify all the individual light sources in the real scene illumination. This means we can only dim or turn off *all* the input lights as one, that may result in residual artifacts if the target lighting levels are very different from those of the input. We handle distant directional light sources – such has light coming through a window – with our view selection method for secondary rays. We use the view forming the closest angle with the ray (see § 4.1.1, 4.2.1), but this is inherently limited by the angular sampling rate of the light source in the input views. We have however, full and independent control over the user-specified new lights.

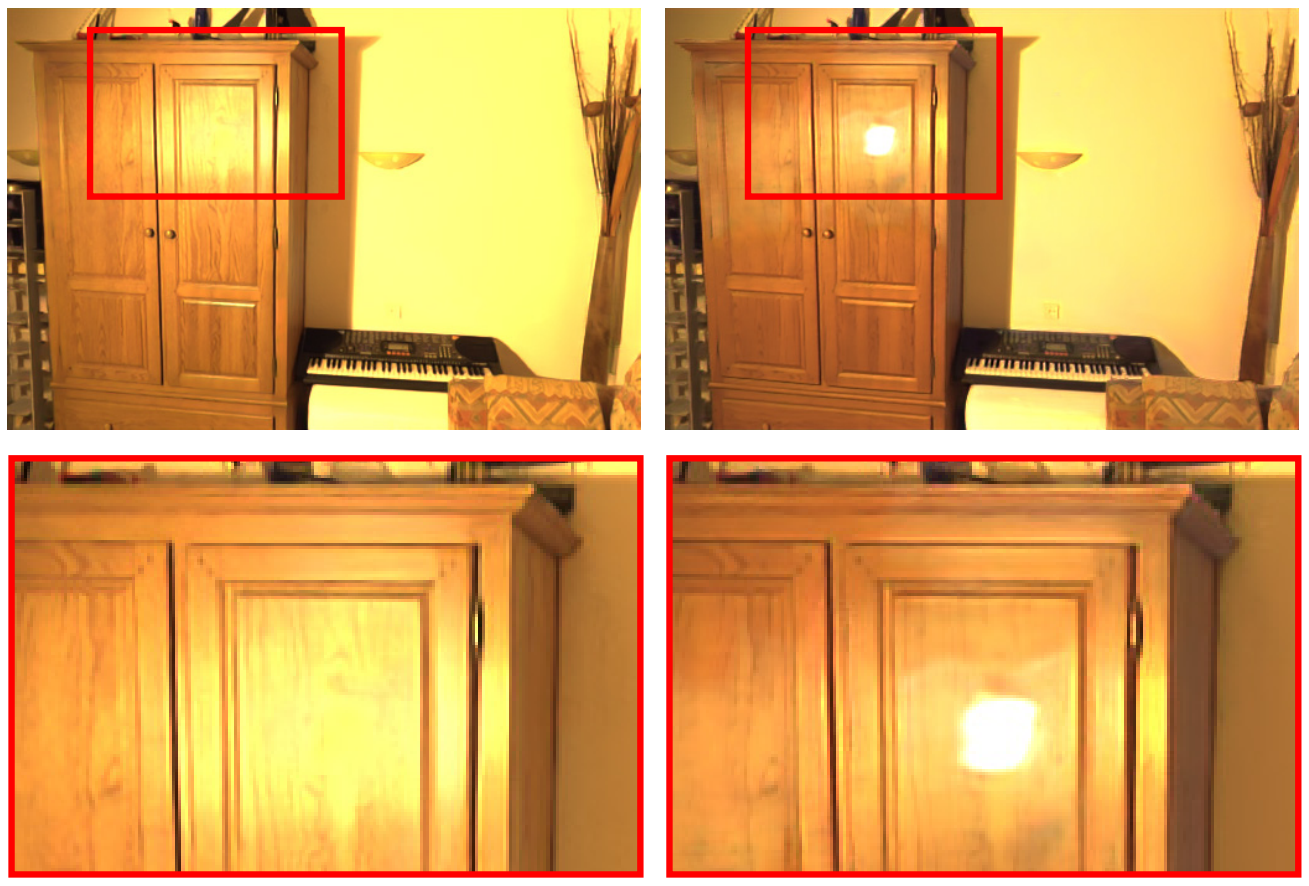

(a) our full solution (b) without the multi-scale module

Fig. 19. **Multi-scale module ablation.** Our multi-scale module provides a wide receptive field that is particularly important to analyze complex materials with a large spatial patterns (a). Without it, the network struggles with rough surfaces for instance, and generates pure-mirror reflections (b) instead of spreading them out properly.

We have tested our method on indoors scenes. Even though our method is general, outdoors scenes would require specific treatment of the sun and sky lighting and specific training data. It may also be necessary to have true high-dynamic range capture since the required dynamic range outdoors is usually higher than indoors, where 14-bit RAW suffices.

Currently, creating animations with moving light sources would require 2-3 minutes of precomputation for each new light position. However, the bulk of this computation is in the added irradiance calculation, which uses path tracing in the space of the input views. This is a relatively simple computation, which traces rays against the simple proxy mesh. With a real-time, hardware-accelerated ray-tracer (e.g., RTX [Burgess 2020]), this step could be made to run at interactive rates, by computing the new irradiance directly in the novel view space. Finally, our simplified illumination model does not hold for anisotropic materials. We expect our model to fall back to a standard blending behavior in such cases.

## 9 CONCLUSION

In this paper, we presented the first method that permits relighting and free-viewpoint navigation in real indoor scenes with glossy materials. Our algorithm combines the strengths of inverse, physically-based and image-based rendering methods. It uses multi-view inputs, a 3D mesh reconstructed by MVS and efficient approximations of physically-based rendering to compute input feature maps from which a convolutional network learns an implicit representation of materials and lighting in the scene. Our feature maps include source and target diffuse irradiance, and mirror images for both the input and novel views. They are reprojected and processed by a neural network to relight and re-render the scene. Our model runs at interactive frame-rates, which enables free-viewpoint navigation. It is trained exclusively on synthetic scenes, yet generalizes to the real world thanks to a data-generation pipeline that accounts for error in the 3D reconstruction step. Our results show controllable and realistic relighting of indoor scenes and plausible renderings of apparent motion for glossy reflections. This opens up a promising avenue for future research, paving the way for general solutions that can handle both indoor and outdoor scenes, and further integrate the flexibility of traditional computer graphics with the ease of capture provided by image-based rendering approaches.

## 10 ACKNOWLEDGMENTS

This research was funded in part by the ERC Advanced grant FUNGRAPH No 788065 (http://fungraph.inria.fr). The authors are grateful to the OPAL infrastructure from Université Côte d'Azur for providing resources and support. The authors thank G. Riegler and J-H. Wu for help with comparisons. Thanks to A. Bousseau and P. Shirley for proofreading earlier drafts. Finally, the authors thank the anonymous reviewers for their valuable feedback.

## A LIGHT-LEVELS ESTIMATION FOR OVEREXPOSED REAL SCENES

In real scenes, the absolute, global light level can be ambiguous because directly visible light sources may saturate the sensor and clip, even in high bit-depth captures. In such cases, we estimate the global energy level using a simple algorithm. We first detect the overexposed regions in 3D space, by filling a sparse voxel grid (2cm effective resolution). When at least 50% of the visible pixels (from the input views) that reproject to the voxel are clipped, we mark the voxel as overexposed. We then cluster the voxels; each clipped voxel initializes a cluster and we iteratively merge clusters whenever their bounding spheres intersect. The 3D region defined by a cluster roughly corresponds to a real-world light source. We compute a source irradiance map $E_i^{src;l}$ for each such light $l$, by directly sampling the defined volume, i.e., we cast rays only towards the light volume and not in the entire hemisphere. When reprojecting the intersection in the input images, we assume constant emittance over the entire detected light region and zero elsewhere. We also compute a separate irradiance map for the non-clipped pixels $E_i^{src-nc}$ as described in § 4.1.1, discarding clipped pixels. We can then model the total irradiance as a linear combination of the non-clipped irradiance map $E_i^{src-nc}$ and per-light irradiance maps $E_i^{src;l}$, with unknown non-negative weights $\alpha_l$:

$$E_i^{src} = E_i^{src-nc} + \sum_l \alpha_l E_i^{src;l} \tag{13}$$

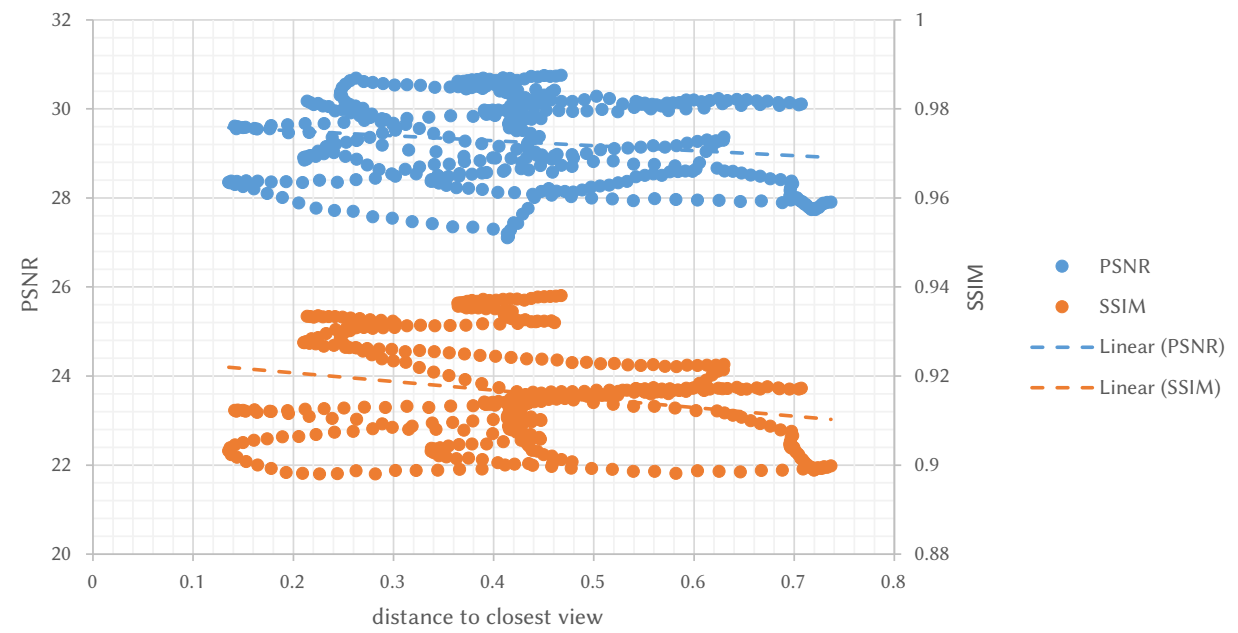

Fig. 20. Graph of error (PSNR and SSIM) for the quantitative evaluation using synthesized viewpoints with respect to the distance to the input cameras; there is no significant correlation between the two. Distance is in meters.

Where $\alpha_l > 0$. We ask the user to click on sets of points $P$ with the same albedo in the scene. We assume the selected points should all have the same albedo $a$, which we approximate as the image value divided by irradiance $\frac{I_i}{E_i^{src}}$. This gives equality constraints to assemble a linear system, such that for $p \in P$:

$$\frac{I_i(p)}{E_i^{src}(p)} = a \tag{14}$$

Leading to:

$$E_i^{src-nc}(p) + \sum_l \alpha_l E_i^{src;l}(p) = \frac{I_i(p)}{a} \tag{15}$$

We solve for the scalar weights $\alpha_l$ using least-squares optimization; these correspond to each light's intensity. When a single light source is clipped, the user only needs to select two points, one lit by the overexposed light and one in its shadow to avoid an ill-conditioned linear system.

## B DATASET STATISTICS FOR THE REAL SCENES

In our quantitative analysis on the synthetic scene *livingRoom2*, we found that the distance between the synthesized viewpoint and input cameras did not significantly correlate with error, as illustrated below (Fig. 20). This shows our model generalizes well to distant viewpoints.

We also show the mean and standard deviation of distance to the input cameras in Table 4.

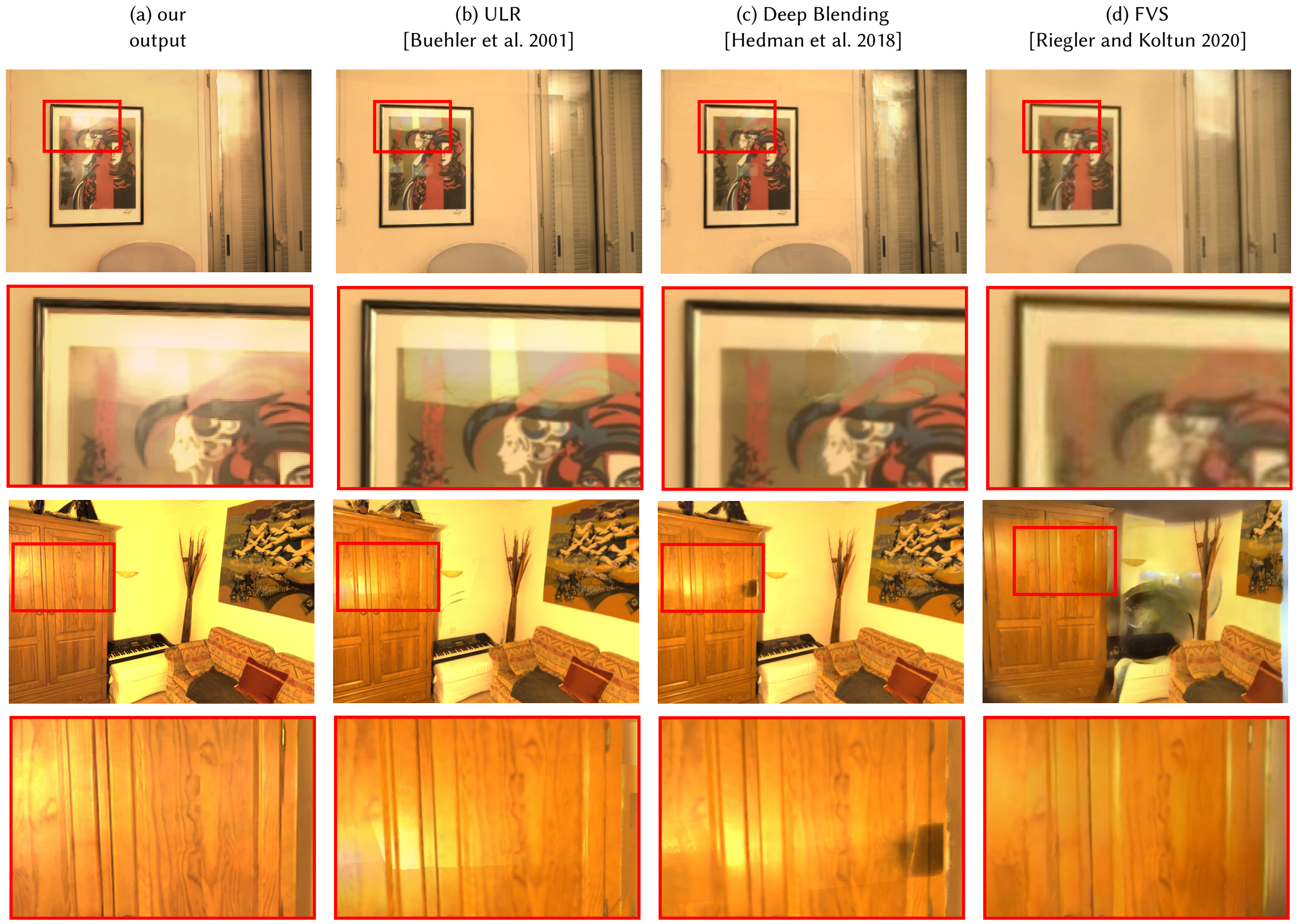

Fig. 21. **View-synthesis comparison.** In the LivingRoom and Sofa scenes, we see that plausible glossy highlights are generated by our method (a) (see also supplemental video for moving highlights). In contrast, ULR (b), Deep Blending (c) and FVS (d) do not faithfully reproduce glossy highlights.

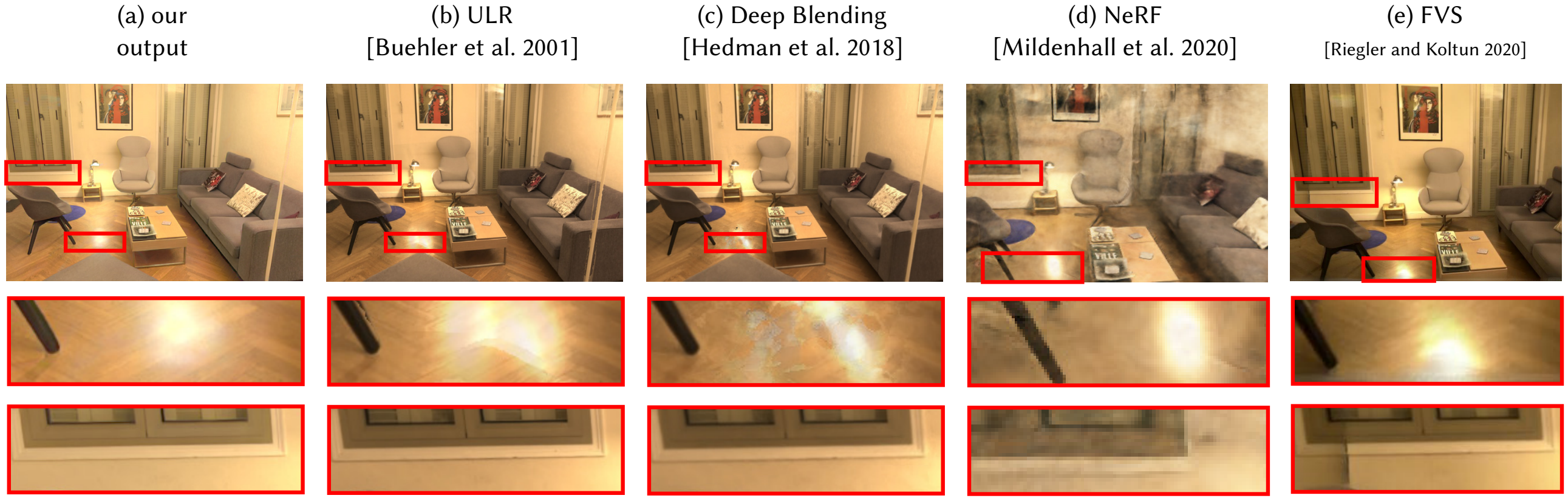

Fig. 22. **View-synthesis comparison.** Our model (a) synthesizes plausible glossy highlights (see supplemental video to better appreciate the effect for moving highlights) whereas ULR (b) and Deep Blending (c) exhibit artifacts like double highlights, cross-fading and missing highlights. NeRF (d) can synthesize plausible highlights, but generally yields much lower quality on large-baseline indoor scenes. FVS (e) suffers from temporal coherence issues and exhibits strong artifacts.

| scene | mean | std |
|---|---|---|
| Living Room | 0.187 | 0.07 |
| Bedroom 1 | 0.174 | 0.11 |
| Bedroom 2 | 0.173 | 0.11 |
| Hall | 0.240 | 0.14 |
| Sofa | 0.192 | 0.10 |
| Kitchen | 0.157 | 0.08 |

Table 4. Mean and standard deviation (std) of distance to input cameras for novel views in the test paths.